\documentclass[10pt,english,floatfix,superscriptaddress,aps,preprint,showpacs]{revtex4}
\usepackage{amsmath}
\usepackage{amssymb}
\usepackage{amsbsy}
\usepackage{amsfonts}
\usepackage{amsopn}
\usepackage{amstext}
\usepackage{graphicx}
\usepackage{amssymb}
\usepackage{amsfonts}
\usepackage{amsmath}
\usepackage{graphicx}
\usepackage[english]{babel}
\usepackage{color}
\usepackage[dvips]{epsfig}
\usepackage[dvips]{graphicx}
\usepackage{float}
\usepackage{units}
\usepackage{textcomp}
\usepackage{babel}

\newcommand\fverb{\setbox\fverbbox=\hbox\bgroup\verb}
\newcommand\fverbdo{\egroup\medskip\noindent%
            \fbox{\unhbox\fverbbox}\ }
\newcommand\fverbit{\egroup\item[\fbox{\unhbox\fverbbox}]}
\newbox\fverbbox

\def\text#1{\mbox{#1}}

\DeclareMathOperator{\e}{e}
\DeclareMathOperator{\sech}{sech}
\DeclareMathOperator{\mx}{max}
\DeclareMathOperator{\mn}{min}
\DeclareMathOperator{\diag}{diag}

\begin{document}
%%%%%%%%%%%%%%%%%%%%%%%%%%%%%%%%%%%%%%%%%%%%%%%%%%%%%%%%%%%%%%%%%%%%%%%%%%%%%%%%%%%%%%%%%%%%%%%%%%%%%%%%%%%%%%%%%%%%%%%%%%%%%%%%%%%%%%%%%%%%%%%%%%%%%%%%%%%%%%%%%%55

\title{Gravitational Kaluza-Klein modes in the String-Cigar Braneworld}
\author{D. F. S. Veras}
\email{franklin@fisica.ufc.br}
\affiliation{Departamento de F\'{i}sica - Universidade Federal do Cear\'{a} (UFC) \\ C.P. 6030, 60455-760
Fortaleza-Cear\'{a}-Brazil}
\author{J. E. G. Silva}
\email{euclides@fisica.ufc.br}
\affiliation{Departamento de F\'{i}sica - Universidade Federal do Cear\'{a} (UFC) \\ C.P. 6030, 60455-760
Fortaleza-Cear\'{a}-Brazil}
\author{W. T. Cruz}
\email{wilami@pq.cnpq.br}
\affiliation{Instituto Federal de Educa\c{c}\~{a}o, Ci\^{e}ncia e Tecnologia do Cear\'a (IFCE) \\ Campus Juazeiro do Norte - CEP 63040-540, Juazeiro do Norte, Cear\'a , Brazil}
\author{C. A. S. Almeida}
\email{carlos@fisica.ufc.br}
\affiliation{Departamento de F\'{i}sica - Universidade Federal do Cear\'{a} (UFC) \\ C.P. 6030, 60455-760
Fortaleza-Cear\'{a}-Brazil}
\date{\today}

\begin{abstract}
In this work the gravitational Kaluza-Klein (KK) modes are analysed on string-like braneworlds in six dimensions,
namely the thin string-like Gherghetta-Shaposhnikov (GS) and the thick string-cigar model.
We find a new massless mode in both modes satisfying the respective Schr\"{o}dinger equation.
By means of numerical analysis for both models, a complete graviton spectrum and eigenfunctions is attained. Besides the linear regime, the KK spectrum exhibits a decreasing behavior with an asymmetric mass gap
with respect to the massless mode. The discontinuity is bigger in the GS model, and it is invariant upon the geometrical flow.
A massive mode is obtained in the GS model with a tiny mass which does not match in any regime.
It turns out that in the string-cigar model, the core brane structure smoothes and amplifies the KK modes near
the brane. Further, the presence of a potential well allows the existence of resonant
massive gravitons for small masses in the string-cigar scenario.
%are studied by means of the resonance method.
\end{abstract}
%%Verificar Abstract%%

\maketitle

%%%%%%%%%%%%%%%%%%%%%%%%%%%%%%%%%%%%%%%%%%%%%%%%%%%%%%%%%%%%%%%%%%%%%%%%%%%%%%%%%%%%%%%%%%%%%%%%%%%%%%%%%%%%%%%%%%%%%%%%%%%%%%%%%%%%%%%%%%%%%%%%%%%%%%%%%%%%%%%%%%%%%%%%

\section{Introduction}
\label{Introduction}

Among the high energy theories proposed in the last years, the braneworld models gained prominence due their fundamental basis on string-theory and since they grant an explanation for the gauge hierarchy and the cosmological constant problems  \cite{RS1, RS2, ADD}.
Once Kaluza-Klein (KK) models \cite{Kaluza, Klein} launched the possibility of extra dimensions, it has been blossomed models where the four-dimensional observable space-time is regarded as a membrane embedded in a higher-dimensional space-time \cite{Akama:1982jy, Rubakov-Shaposhnikov, Rubakov-Shaposhnikov-2}.  The Randall-Sundrum (RS) model enhanced this hypothesis by assuming an infinite extra dimension  whose warped geometry yields to a compact transverse space \cite{RS2}. In the sequel, some models appeared proposing the brane be generated by a topological defect. In five dimensions, domain walls have been used to represent the brane \cite{Rubakov-Shaposhnikov, Visser, Gremm}, whilst in six dimensions, with axial symmetry the geometry resemblances that of the cosmic strings \cite{Israel}. The resulting scenario is the so-called string-like braneworld \cite{Navarro}.

Systems composed by complex scalar field (global) and gauge field (local) often describe string-like objects, such as the vortices and the cosmic strings \cite{Gregory1987}. However, unlike the domain walls, there is no complete (interior and exterior) string-like solution analytically known, even in flat space-time. In warped braneworlds, Cohen-Kaplan \cite{CohenKaplan-1999}, Gregory \cite{Gregory-2000} and Olaganesi-Vilenkin \cite{Olasagasti:2000gx} studied the exterior solution of a global string-like brane with and without bulk cosmological constant. The local case was addressed numerically by Giovaninni \textit{et al} which acquired a smooth geometry satisfying the dominant energy condition \cite{GMS}.

Apart from the particular model used to compose the brane core, the exterior warped string-like solution with cosmological constant has a conformally-flat behavior that extends the five-dimensional Randall-Sundrum model \cite{Ponton:2000gi}. In the infinitely thin core brane limit, the resulting model is called the Gherghetta-Shaposhnikov (GS) model \cite{GS}. The GS model provides a correction to the Newtonian potential due to the KK modes less than in the RS model \cite{GS}. Another advantage over the RS model is that the Kaluza-Klein massless mode of the gauge field is naturally
localized on the string-brane without any further coupling \cite{Oda-1,Oda-2}. The fermionic field, in its turn, can be trapped in the GS brane through a gauge minimal coupling only \cite{Oda-1, Oda-2, Liu-Zhao}.

Besides the interesting features for the fields, the string-like braneworlds also exhibit a rich geometric structure. In fact, the two extra dimensions form a transverse manifold with its internal symmetries and properties that reflect on the brane tension and geometry \cite{Navarro}. The transverse manifold used in the GS model is a disc whereas Kehagias proposed a conical space to provide an explanation for the cosmological constant problem \cite{Kehagias:2004fb}.
Garriga-Porrati studied the effects generated by a football-shaped manifold
\cite{Garriga:2004tq} and Gogberashvili \textit{et al} addressed the fermion generations
problem by means of an apple-shaped space \cite{Gogberashvili:2007gg}. In a supersymmetry setup, de Carlos and Moreno found a gravity trapping solution
without cosmological constant and that asymptotically has a constant transverse radius, named as a cigar-like universe \cite{deCarlos:2003nq}.

%The idea of extra dimensions in physics is not new. It  dates from the twenties with the , which propose a theory of unification for the electromagnetic and gravitational forces, extending the general theory of relativity to a 5D space-time. In braneworld models the observable universe is represented by a . In the mid-1980s, there were proposals that this membrane was generated by a topological defect .

%As regards to the study of field localization in string-like branes, the gravitational, matter, bosonic and fermionic fields were treated . In particular, the Gherghetta-Shaposhnikov (GS) model \cite{GS}, localizes gravity in an infinite thin string-like brane embedded in a six dimensional static and axisymmetric bulk built from a warped product between the brane and a disc.

In a recent work published by two of the present authors \cite{Charuto}, the so-called cigar soliton is used to construct an interesting smooth interior and exterior string-like geometry referred as the string-cigar. The cigar soliton
is a two-dimensional self-similar solution of the Ricci flow, a geometric flux driven by the
Ricci tensor \cite{Chow,Hamilton}. There are important applications of the Ricci flow in differents
branches of Physics as in the sigma models \cite{Friedan:1980jf}, Euclidean black holes \cite{Headrick:2006ti},
topological massive gravity \cite{Lashkari:2010iy}, ADM mass \cite{xianzhe} and in the Heisenberg
model in statistical mechanics \cite{Orth:2012ri}. The Ricci flow defines a family of geometries developing under an evolution parameter $c$. Therefore,
the string-cigar scenario proves the changes of the brane properties due to the
geometric flow \cite{Silva:2011yk}.
Since the Ricci soliton extends the Einstein manifolds, the string-cigar
geometry can be realised as an augmented codimension two model proposed by Randjbar-Daemi
and Shaposhnikov whose transverse space is a Ricci flat manifold \cite{RandjbarDaemi:2000ft}.
%It turned out that the brane source satisfies the dominant energy condition and it undergoes changes in the brane tensions and in the brane gravitational constant \cite{Charuto}.

Another smoothed thick string-like model which leads to a geometrical flow in the
transverse space was built with a section of the resolved conifold; an important orbifold in
string theory \cite{Silva:2011yk}. Both models yield to
regular geometries that asymptotically recover the GS model \cite{Charuto}. Nevertheless, the
string-cigar model satisfies all the regularity conditions required to ensure a well-behaved
3-brane at the origin whereas the resolution parameter plays the
role of the radius of the fifth dimension what violates the conical behavior near the origin \cite{Silva:2011yk}.

The string-cigar scenario enables the existence of a localized gravitational massless-mode
which effectively describes the gravity on the brane \cite{Charuto}.
The massless (or zero) mode in this scenario has the same asymptotically exponential
behavior of that in the GS model, but near the origin it is smooth and has a bell-shape.
Hence, the string-cigar scenario smoothes the geometry and the zero-mode near the origin.
As the geometry undergoes the Ricci flow in the transverse space, the massless mode changes
its width and amplitude \cite{Charuto}. In this work, we find a new massless mode for the GS and the string-cigar models
by the Schr\"{o}dinger approach which exhibits a more localized behavior.

Another noteworthy feature of the string-cigar model is its inhomogeneous source \cite{Charuto}.
Indeed, the maximum of the stress-energy tensor components is displaced from the origin what
suggests the brane core is shifted \cite{Charuto}. Giovaninni \textit{et al} found a similar behavior for an Abelian vortex with higher winding number \cite{GMS}. The gravitational
massless mode shares the same profile, and it shows the influence of the geometric changes on the
physics of the brane \cite{Charuto}. Moreover, as pointed out by Tinyakov and Zuleta, the source for the GS model does not satisfy the dominant energy condition \cite{Tinyakov:2001jt}. On the other hand, the string-cigar source undergoes a phase transition whereupon some configurations fulfill all the energy conditions \cite{Charuto}. Further, as higher the bulk cosmological constant closer to the GS model is the string-cigar source.

Nonetheless, the complexity of the differential equation for the KK modes turns
the numerical analysis the most subtle approach to obtain the whole spectrum of masses and
eigenfunctions \cite{Gravity-Quasi-Ress}. Thus, the foremost purpose of this work is to attain the massive KK spectrum and
to analyse how it behaves upon the Ricci flow
that the cigar undergoes. 

Although Gherghetta and Shaposhnikov have found a complete set
of eigenfunctions for their thin-string model, the corresponding KK masses expression for the discrete states is valid only for large values of a discretization index $n$ \cite{GS}. In this limit, the masses increase linearly but for small $n$ the behavior
is yet unknown. Then, our numerical analysis provides the full KK mass spectrum of the GS and string-cigar scenarios. Both models show a decreasing behavior for small $n$ with a mass gap between the two
patterns. Moreover, the GS model presents a massive mode with tiny mass and amplitude which is absent in the string-cigar model. The parallel between the two 
models provides more information of how the fields behave on singular and smooth scenarios in
six dimensions. Giovaninni \textit{et al} have numerically studied the interior and exterior
 geometry of a string-like braneworld, but have not concerned with the KK spectrum \cite{GMS}.
In this article, numerical methods are also used to find and to study gravitational resonant
massive modes on both GS and string-cigar models. Even though the analogous Schr\"{o}edinger potential
exhibits an infinite potential well around the origin for the string-cigar model,
resonant modes are allowed in contrast with the GS model where they are absent. It turns out that
the resonant modes appear only for small masses.

%The choice of the Hamilton cigar as the transverse space provides inside and outside solutions enabling to study corrections of GS model due the effects near and  inside the core of the string. The Hamilton cigar is a two dimensional regular manifold, conformal to the disc and asymptotically flat \cite{Chow}.

%In this paper, we use suitable numerical methods to study the gravitational Kaluza-Klein (KK)
%modes of the thin string-like GS and thick String-Cigar models.

This paper is organized as follows: in the section \ref{Review} the main characteristics and results of the GS and string-cigar models are reviewed. Next, in section \ref{NumericalResults-1}, the numerical results which concern to the spectrum and eigenfunction are presented. In section \ref{NumericalResults-2}, the study of KK massive modes as resonant states is developed. Finally, in the section \ref{Conclusions}, some final remarks and perspectives are outlined.

%%%%%%%%%%%%%%%%%%%%%%%%%%%%%%%%%%%%%%%%%%%%%%%%%%%%%%%%%%%%%%%%%%%%%%%%%%%%%%%%%%%%%%%%%%%%%%%%%%%%%%%%%%%%
%%%%%%%%%%%%%%%%%%%%%%%%%%%%%%%%%%%%%  REVIEW %%%%%%%%%%%%% %%%%%%%%%%%%%%%%%%%%%%%%%%%%%%%%%%%%%%%%%%%%%%%%
%%%%%%%%%%%%%%%%%%%%%%%%%%%%%%%%%%%%%%%%%%%%%%%%%%%%%%%%%%%%%%%%%%%%%%%%%%%%%%%%%%%%%%%%%%%%%%%%%%%%%%%%%%%%

\section{String-like Braneworlds}
\label{Review}

%In this section we present the definitions and the main features of the string-like models we concerned in this work.

Consider a six dimensional space-time $\mathcal{M}_{6}$ built from the warped product between a 4D Lorentzian manifold $\mathcal{M}_{4}$ and the 2D
Riemannian manifold $\mathcal{M}_{2}$. Hereinafter, we refer $\mathcal{M}_{4}$ as a 3-brane and $\mathcal{M}_{2}$ as the transverse space.

A string-like braneworld is a static 6D $\mathcal{M}_{6}$ endowed with an axial symmetry in the transverse space. A subtle metric for this model is given by
\cite{Navarro,CohenKaplan-1999,Gregory-2000,Olasagasti:2000gx,GMS,GS,Oda-1}
\begin{equation}
ds_{6}^{2} =g_{AB}(x,\rho,\theta)dx^{A}dx^{B}= \sigma(\rho)g_{\mu\nu}(x)dx^{\mu}dx^{\nu} - d\rho^{2} - \gamma(\rho)d\theta^{2},
\label{MetricaGS}
\end{equation}
where $x$ are brane coordinates, $\rho \in [0,\infty)$ and $\theta \in [0, 2\pi]$ are the extra dimensions and $\sigma$ and $\gamma$ are the so-called warp factors.

The additional regularity conditions, 
\begin{eqnarray}
\sigma(0) = 1  & ,  &  \sigma'(0) = 1,\nonumber\\
\gamma(0) = 0  & ,  &  \left(\sqrt{\gamma(0)}\right)'=1,
\label{RegularityConditions}
\end{eqnarray}
are imposed in order to avoid singularities \cite{Navarro}, where the ($'$) stands for the derivative $\partial_{\rho}$. The conditions for $\sigma$ in Eq. (\ref{RegularityConditions}) are already present in the RS models \cite{RS1,RS2} whereas the
further assumption for $\gamma$ reflects the smooth but conical behavior near the origin \cite{Navarro,GMS,GS,Tinyakov:2001jt}.

%\begin{equation}
%\begin{split}
%\sigma(0) = 1  & ,\hspace{1cm}  \sigma'(0) = 1,\\
%\gamma(0) = 0  & ,\hspace{1cm}  (\sqrt{\gamma(0)})'=1.
%\end{split}
%\end{equation}

Regardless the particular model for the source, an axisymmetric stress-energy tensor of the following form is adopted \cite{Navarro,GMS,GS}
\begin{equation}
T^{A}_{B}=\diag(t_{0},t_{0},t_{0},t_{0},t_{\rho},t_{\theta}).
\end{equation}
For a global string, for instance, $t_{\rho}=-t_{\theta}$ \cite{CohenKaplan-1999,Gregory-2000,Olasagasti:2000gx}. In the presence of a bulk cosmological
constant $\Lambda$ and for a flat brane $\mathcal{M}_{4}$, the Einstein equation reads
 \begin{equation}
\label{Einstein}
 R_{ab}-\frac{R}{2}g_{ab}=-\kappa_{6}(\Lambda g_{ab}+T_{ab}),
\end{equation}
that for the metric ansatz (\ref{MetricaGS}), yields to
\begin{eqnarray}
 \frac{3}{2}\left(\frac{\sigma'}{\sigma}\right)' + \frac{3}{2}\left(\frac{\sigma'}{\sigma}\right)^{2} + \frac{3}{4}\frac{\sigma'}{\sigma}\frac{\gamma'}{\gamma} + \frac{1}{4}\left(\frac{\gamma'}{\gamma}
\right)^ { 2 } +\frac {1} {2} \left(\frac { \gamma'}{\gamma}\right)' &  =   &   -\kappa_{6}(\Lambda+t_{0}(\rho)),\\
\frac{3}{2}\left(\frac{\sigma'}{\sigma}\right)^{2}+\frac{\sigma'}{\sigma}\frac{\gamma'}{\gamma} &   =   &   -\kappa_{6}(\Lambda+t_{\rho}(\rho)), \\
2\left(\frac{\sigma'}{\sigma}\right)'  +  \frac{5}{2}\left(\frac{\sigma'}{\sigma}\right)^{2}    &   =   &    -\kappa_{6}(\Lambda+t_{\theta}(\rho)),
\end{eqnarray}
where $\kappa_{6}$ is the six-dimensional gravitational constant related to the six-dimensional energy scale by $\kappa_{6}=\frac{8\pi}{M_{6}^{4}}$ \cite{Charuto}.

Analysing the vacuum configuration, the equation %the Einstein equations
%(\ref{Einstein})
\begin{eqnarray}
\label{angulareinsteinequation}
2\left(\frac{\sigma'}{\sigma}\right)'  +  \frac{5}{2}\left(\frac{\sigma'}{\sigma}\right)^{2}    &   =   &    -\kappa_{6}\Lambda
\end{eqnarray}
determines the warp function. Defining
\begin{equation}
y(\rho)=\frac{\sigma'}{\sigma},
\end{equation}
the Eq. (\ref{angulareinsteinequation}) turns to be
\begin{equation}
y'+\frac{5}{4}y^{2}-\frac{\kappa_{6}|\Lambda|}{2}=0,
\end{equation}
whose solution is
\begin{equation}
\label{vacuumwarpfactor}
y(\rho)=c\tanh{\frac{5c}{4}(\rho + \rho_0)},
\end{equation}
where
\begin{equation}
c^2 = -\dfrac{2\kappa_6}{5}\Lambda,
\end{equation}
and $\rho_0$ is an integration constant. Integrating Eq. (\ref{vacuumwarpfactor}) yields to the warp factor
\begin{equation}
\label{divergentwarpfactor}
\sigma(\rho)=\sigma_{0}\cosh^{\frac{4}{5}}{\left(\frac{5}{4}c \rho\right)},
\end{equation} 
with a integration constant $\sigma_{0}$. The warp factor in (\ref{divergentwarpfactor}) provides an infinite volume to the transverse space which leads to a non four-dimensional effective
gravitational theory.

The relation between the bulk Planck mass $M_{6}$ and brane Planck masses $M_{4}$ is given by
\begin{eqnarray}
M^{2}_{4} & = & 2\pi M_{6}^{4}\int_{0}^{\infty}{\sigma(\rho)\sqrt{\gamma(\rho)}d\rho}.
\label{planckmass}
\end{eqnarray}
Then, for a string-like model with warp function (\ref{divergentwarpfactor}) the brane Planck scale diverges.

%\textit{\textbf{Acho que este parágrafo abaixo pode ser suprimido tendo em vista que não foi calculado tomando os parâmetros desse paper. Se ficar precisa definir o que significam as letras que aparecem.}
%}

Another important characteristic of the string-like models are the string tensions $\mu_i$ defined by \cite{GS,GMS}
\begin{eqnarray}
 \mu_{i}(c) & = & \int_{0}^{\epsilon}{t_{i}(\rho,c)\sigma^{2}(\rho,c)\sqrt{\gamma(\rho,c)}d\rho},
\end{eqnarray}
which determines the matching between an internal and an external geometric solution made in the boundary of the core with width $\epsilon$ \cite{Navarro,GS,GMS}.
%\subsection{String-like Braneworlds}

Once presented the general aspects of the string-like scenario, the behavior of the small metric fluctuations around this configuration is  given below.

Performing the following conformal invariant perturbation \cite{GS,GMS,Charuto,Csaki:2000fc}
\begin{equation}
ds^{2}_{6}=(\sigma(\rho)\eta_{\mu\nu}+h_{\mu\nu}(x,\rho))dx^{\mu}dx^{\nu}+d\rho^{2}+\gamma(\rho) d\theta^{2},
\end{equation}
the first-order perturbed Einstein equation (\ref{Einstein}) yields to \cite{GS,GMS,Charuto,Csaki:2000fc}
\begin{equation}
\label{pertubationequation}
\Box_{6} h_{\mu\nu}=\partial_{A}(\sqrt{-g_{6}}\eta^{AB}\partial_{B}h_{\mu\nu})=0.
\end{equation}
Thus, the tensorial perturbation $h_{\mu\nu}$ can be regarded as a tensorial field (graviton) propagating in the bulk.

Assuming the usual Kaluza-Klein decomposition \cite{RS2,GS,Oda-1,Charuto}
\begin{equation}
\label{kkdecomposition}
h_{\mu\nu}(x, \rho, \theta) = \sum_{l,m=0}^{\infty}\phi_{m,l}(\rho)\e^{il\theta}\hat{h}_{\mu\nu}(x),
\end{equation}
and a free-wave dependence on the 3-brane
\begin{equation}
\Box_{4}h_{\mu\nu}(x^{\xi}) = m_0^2 h_{\mu\nu}(x^{\xi}),
\end{equation}
the graviton equation of motion (\ref{pertubationequation}), for a conformal scenario where $\gamma(\rho)=\sigma(\rho) \beta(\rho)$, reads \cite{GS,Charuto}
\begin{equation}
\left(\sigma^{\frac{5}{2}}\sqrt{\beta}\phi_{m,l}^{\prime}(\rho)\right)^{\prime} + \sigma^{\frac{3}{2}}\sqrt{\beta}\left(m_0^{2} - \frac{l^{2}}{\beta^{2}}\right)\phi_{m,l}(\rho) = 0.
\label{Sturm-Liouville-Charuto}
\end{equation}
The function $\beta$ is responsible by the conical behavior \cite{Silva:2011yk,Charuto}. The Eq. (\ref{Sturm-Liouville-Charuto}) describes the radial behavior of the graviton on
string-like scenarios. The presence of the angular number $l$ turns the spectrum degenerated \cite{GS}.
In addition, due to the axial symmetry, the boundary conditions are \cite{GS,GMS,Charuto}
\begin{equation}
\phi^{\prime}(0) = \phi^{\prime}(\infty) = 0.
\label{Cond-Cont-GS}
\end{equation}
The radial equation (\ref{Sturm-Liouville-Charuto}) and the boundary conditions (\ref{Cond-Cont-GS}) provide a set of solutions whose orthogonality relation is given by
\begin{equation}
\int_0^{\infty}d\rho \hspace{0.07cm} \sigma^{\frac{3}{4}}\hspace{0.07cm}\sqrt{\beta}\hspace{0.07cm}\phi_{m,l}\phi_{n,l'} = \delta_{mn}\delta_{ll'}.
\label{GS-OrthonormalRelation}
\end{equation}

The eigenvalues of Eq. (\ref{Sturm-Liouville-Charuto}) satisfying the boundary conditions (\ref{Cond-Cont-GS}) are called the KK spectrum (mass) and the respective
eigenfunctions are called the KK states. Among the KK states there is one for a vanishing mass, called massless or zero mode. From Eq. (\ref{Sturm-Liouville-Charuto}), the massless mode has the form
\begin{equation}
\label{slmassless}
\phi_{0}(\rho)=A_{1}\int_{0}^{\rho}{\sigma^{-\frac{5}{2}}\beta^{-\frac{1}{2}}d\rho'} + A_{2},
\end{equation}
where $A_1$ and $A_2$ are constants. A similar massless mode was found by Csaki \textit{et al} for non-conformally flat space-times \cite{Csaki:2000fc}.

A suitable way to study the KK spectrum consists of turn Eq. (\ref{Sturm-Liouville-Charuto}) into a Schr\"{o}dinger-like equation. By taking the change of independent variable
\begin{equation}
z(\rho) = \int^{\rho} \sigma^{-1/2}d\rho^{\prime}
\label{Z-Transformation}
\end{equation}
and of a dependent variable
\begin{equation}
\phi_m(z) = u(z)\Psi_m(z),
\label{Chi-u-Psi}
\end{equation}
where
\begin{equation}
\dfrac{\dot{u}}{u} + \dfrac{\dot{\sigma}}{\sigma} + \dfrac{1}{4}\dfrac{\dot{\beta}}{\beta} =0,
\label{Funcao-u}
\end{equation}
the radial equation (\ref{Sturm-Liouville-Charuto}) yields to
\begin{equation}
-\ddot{\Psi}_m(z) + U(z)\Psi_m(z) = m^2\Psi_m(z),
\label{EqSchroedinger}
\end{equation}
with
\begin{equation}
U(z) = \dfrac{\ddot{\sigma}}{\sigma} + \dfrac{1}{2}\dfrac{\dot{\sigma}}{\sigma}\dfrac{\dot{\beta}}{\beta} - \dfrac{3}{16}\left(\dfrac{\dot{\beta}}{\beta}\right)^2 + \dfrac{1}{4}\dfrac{\ddot{\beta}}{\beta} + \dfrac{l^2}{\beta}.
\label{Potencial}
\end{equation}

The boundary conditions (\ref{Cond-Cont-GS}) implies the following boundary condition for $\Psi(z)$:
\begin{equation}
\label{schrodingerboundarycondition}
\begin{split}
u^{\prime}(0)\Psi(0) + u(0)\Psi^{\prime}(0) & = 0,\\
u^{\prime}(\infty)\Psi(\infty) + u(\infty)\Psi^{\prime}(\infty) &= 0.
\end{split}
\end{equation}

Besides the bijective relation between the Sturm-Liouville and the Schr\"{o}dinger approaches, the last provides information about resonant modes that we  consider in section (\ref{NumericalResults-2}). In the follows, the geometrical features and the properties of Eq. (\ref{Sturm-Liouville-Charuto}) in the GS model and the string-cigar model are studied.

%=================================================================================
%==========================| The GS Model |=======================================
%=================================================================================

\subsection{The GS model}

Gherghetta and Shaposhnikov found a vacuum solution of Einstein equations given in Eq. (\ref{Einstein}) that localizes the gravity on the string-like brane.  Assuming that \cite{GS}
\begin{equation}
\frac{\sigma'}{\sigma}=-c,
\end{equation}
which can be obtained from the hyperbolic tangent function in Eq. (\ref{vacuumwarpfactor}) in one of its asymptotic values, the Eq. (\ref{angulareinsteinequation}) yields to
the following warp function
\begin{equation}
\sigma(\rho) = \e^{-c\rho}.
\label{GS-WarpFactor}
\end{equation}
Moreover, for
\begin{equation}
\beta(\rho) = R_0^2,
\label{GS-AngularFactor}
\end{equation}
the GS model describes a $AdS_{6}$ space-time \cite{GS}. Since the GS model is built from the vacuum solution, it can be regarded as the space-time of a thin string-like braneworld, i.e., $\epsilon \rightarrow 0$ \cite{GMS,Tinyakov:2001jt}. In addition, the GS solution does not satisfy the regularity conditions at the origin (\ref{RegularityConditions}).

In the GS model, the graviton obeys the radial equation \cite{GS}
\begin{equation}
\phi_m^{\prime\prime} - \dfrac{5}{2}c\phi_m^{\prime} +\left(m_0^2 - l^2/R_0^2\right)\e^{c\rho}\phi_m = 0,
\label{Sturm-Liouville-GS}
\end{equation}
which general solution, can be written in terms of the Bessel functions
\begin{equation}
\phi_m(\rho) = \e^{\frac{5}{4} c\rho}\left[ B_1 J_{5/2}\left(\frac{2m}{c}\e^{\frac{1}{2}c\rho}\right) + B_2 Y_{5/2}\left(\frac{2m}{c}\e^{\frac{1}{2}c\rho}\right)\right],
\label{GS-MassiveModes}
\end{equation}
where $B_1$ and $B_2$ are arbitrary constants and $m = m_0^2 - l^2/R_0^2$. This solution grows exponentially revealing that massive modes are not localized on the brane \cite{GS}.
%\subsubsection{Massless Mode}

From Eq. (\ref{slmassless}), the GS massless mode has the form
\begin{equation}
\phi_{0}(\rho) = A_{1}\e^{\frac{5}{2}c\rho}+A_{2}.
\label{GSzeromode}
\end{equation}
Among the two solutions in Eq. (\ref{GSzeromode}), only the $\phi_{0} = A_{2}$ satisfies the orthogonality relation (\ref{GS-OrthonormalRelation}) \cite{GS}.
Gherghetta and Shaposhnikov defined an orthonormal solution by \cite{GS}
\begin{equation}
\psi_m(\rho) = \e^{-\frac{3}{4} c\rho}\phi_m(\rho),
\label{GS-WaveFunctionFlat}
\end{equation}
so that, the zero-mode becomes
\begin{equation}
\psi_0(\rho) = \sqrt{\dfrac{3c}{2R_0}}\e^{-\frac{3}{4} c \rho}.
\label{Modo-Zero-GS}
\end{equation}
%It has a compact support near origin showing that massless gravitons are trapped on the brane \cite{GS}.
Therefore, a massless mode is localized in the thin-string brane \cite{GS}.
However, this solution does not satisfy the boundary conditions (\ref{Cond-Cont-GS}) at $\rho=0$, because the warp factor (\ref{GS-WarpFactor}) does not obey the usual regularity conditions. On the other hand, by means of the Schr\"{o}dinger approach, another more localized massless
mode was found. Indeed, the Schr\"{o}dinger equation (\ref{EqSchroedinger}) for the GS model is given by
\begin{equation}
 -\ddot{\Psi}_m + \frac{6}{z^{2}}\Psi_m = m^{2}\Psi_m,
\label{GSSE}
\end{equation}
where $z=\frac{2}{c}\e^{\frac{c}{2}\rho}$ and by Eq. (\ref{Funcao-u}), the relation between $\phi_m$
and $\Psi_m$ is
\begin{equation}
 \phi_m = C_0\e^{c\rho}\Psi_m.
 \label{GSrelation}
\end{equation}

For $m=0$, Eq. (\ref{GSSE}) has the solutions
\begin{eqnarray}
 \Psi_{0} &	=	&	C_{1}z^{3} + C_{2}z^{-2}\nonumber\\
  &	=	&	C_{1}\e^{\frac{3}{2}c\rho} + C_{2}\e^{-c\rho},
\end{eqnarray}
which using Eq. (\ref{GSrelation}) yields to the zero-mode (\ref{GSzeromode}). In order to $\Psi_{0}$
be normalized, we set $C_{1}=0$. Therefore, a localized zero-mode satisfying the analogous
Schr\"{o}dinger equation in the GS model is given by
\begin{equation}
\Psi_{0}(\rho)= \sqrt{\frac{7c}{2R_0}}\e^{-c\rho}.
\label{GSnewzeromode}
\end{equation}
Thus, the zero-mode $\Psi_0$ in Eq. (\ref{GSnewzeromode}) has a bigger value at the origin and a bigger
decay rate compared with $\psi_0$ proposed in Ref. \cite{GS}. Further, the massless mode (\ref{GSnewzeromode}) satisfies the boundary conditions (\ref{schrodingerboundarycondition}).

For $m\neq 0$, the solution of GS schrodinger equation is
\begin{eqnarray}
 \Psi_m(z)=\sqrt{\frac{2}{m\pi}}\left(\frac{1}{mz}\right)^{2}\Big[(m^{2}z^{2}-3mz-3)(\cos(mz)-\sin(mz))\Big].
\label{GSmassivemodeschrodinger}
\end{eqnarray}
Since $\Psi_m(z)$ is not defined for $m=0$, we can not obtain the massless mode (\ref{GSnewzeromode})
from Eq. (\ref{GSmassivemodeschrodinger}). This shows a mass gap in GS model which we explore
numerically in section (\ref{NumericalResults-1}).
Moreover, in the asymptotic limit the massive modes $\Psi_{m}$ assume the plane wave form given by
\begin{equation}
\label{GSasymptotic}
\Psi_{m}(z\rightarrow \infty)=\sqrt{\frac{2}{m\pi}}(\cos (m x) - \sin (m x)).
\end{equation}

%\subsubsection{Espectrum and Massive Modes}
%The massive modes (solutions for $m^2 > 0$ of the equation (\ref{Sturm-Liouville-GS})) is obtained

The KK spectrum can be obtained by inserting a finite radial distance cutoff
$\rho_{\mx}$ and imposing the boundary conditions (\ref{Cond-Cont-GS}). Then, the spectrum is
obtained by the zeroes of the Bessel function
\begin{equation}
J_{\frac{3}{2}}\left(\frac{2m_{n}}{2}\e^{\frac{c}{2}\rho_{\mx}}\right)=0
\end{equation}
that for large index $n$ yields to \cite{GS}
\begin{equation}
m_n \simeq c\left( n - \frac{1}{2} \right)\frac{\pi}{2}\e^{-\frac{c}{2}\rho_{\mx}}.
\label{GS-Spectrum}
\end{equation}
The linear relation between the mass and the index $n$ is rather important in the Newtonian potential correction \cite{GS}. Nevertheless, the formula (\ref{GS-Spectrum}) is valid only for great $n$. In section (\ref{NumericalResults-1}), the eigenfunctions of the GS model and a complete spectrum $m_n$ is numerically obtained.

%=================================================================================
%=======================| The String-Cigar Model |================================
%=================================================================================

\subsection{The string-cigar model}

An extension to GS model, the so-called string-cigar braneworld \cite{Charuto}, is built from a warped product between $3$-brane and the cigar soliton space. The cigar soliton is a two dimensional stationary solution for the Ricci flow
\begin{equation}
\dfrac{\partial}{\partial \lambda} g_{ab}(\lambda) = - 2R_{ab}(\lambda),
\end{equation}
where $\lambda$ is a metric parameter and $R_{ab}$ is the Ricci tensor \cite{Chow}. An axisymmetric metric for the cigar soliton reads as \cite{Hamilton}
\begin{equation}
ds_{\lambda}^{2} = d\rho^{2} + \frac{1}{\lambda^2}\tanh^{2}(\lambda\rho) \hspace{0.1cm} d\theta^{2}.
\end{equation}
It is straightforward to see that the cigar soliton is a smooth manifold.

The main idea of the string-cigar scenario is to use the cigar soliton as the transverse space in order to smooth the GS model. The Ricci flow defines a family of string-like branes and the evolution of the transverse space is related to variations on the geometric and physical quantities \cite{Charuto}. Since the asymptotic value of the scalar curvature depends on the evolution
parameter \cite{Charuto}, the geometric flow represents a variation of the bulk cosmological constant. Then, $\lambda$ and $c$ are equivalent labels. The string-cigar scenario is asymptotically flat as the disc of radius $1/\lambda = R_{0}$ used in the GS model \cite{Chow}. However, near the origin, the $\tanh^2\rho$ term smoothes the geometry and provides a thickness to the brane \cite{Charuto}.

%Indeed, the metric factors $\sigma(\rho,c)$ and $\gamma(\rho,c)$ are parameterized by $c$
%and this defines a family of string-like geometries undergoing a geometrical flow \cite{Charuto}.

The warp factor and the angular metric component proposed are, respectively \cite{Charuto},
\begin{equation}
\sigma(\rho,c) = \e^{-(c\rho - \tanh (c\rho))}
\label{Charuto-WarpFactor}
\end{equation}
and
\begin{equation}
\gamma(\rho,c) = \dfrac{1}{c^2}\tanh^2(c\rho) \sigma(\rho,c).
\label{Charuto-AngularFactor}
\end{equation}

The metric (\ref{MetricaGS}) with equations (\ref{Charuto-WarpFactor}) and (\ref{Charuto-AngularFactor}) represents a space-time inside and outside a string-like defect which satisfies all the regularity conditions given in Eq. (\ref{RegularityConditions}) \cite{Charuto}.

\begin{figure}[h]
 \centering
    \includegraphics[width=0.65\textwidth]{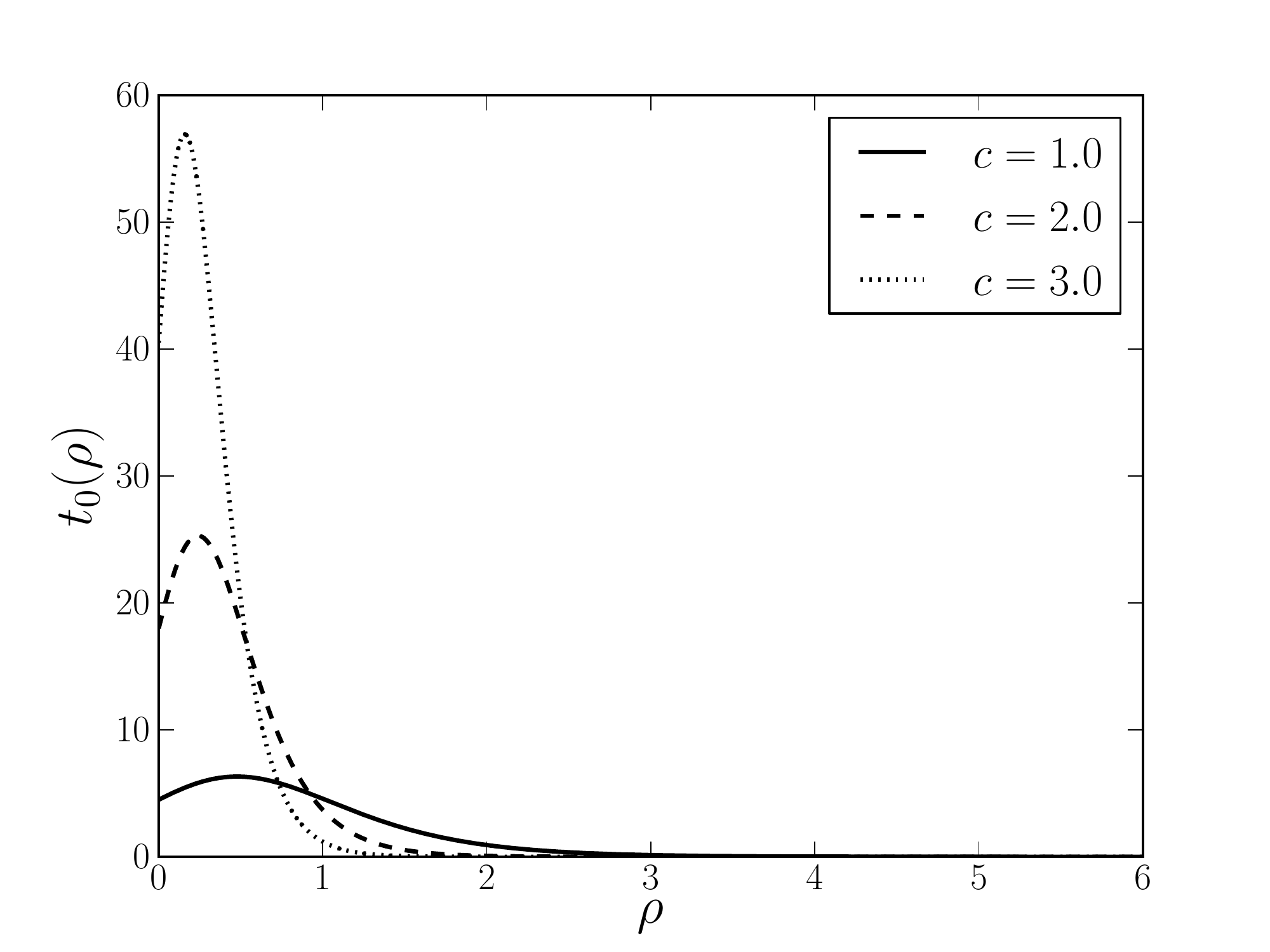}
 \caption{Energy density of the string-cigar braneworld.}
 \label{Fig-DensidadeEnergia}
\end{figure}

The Einstein equation provides the energy density \cite{Charuto}:
\begin{equation}
t_0(\rho,c) = \dfrac{c^2}{\kappa_6}\left(7\sech^{2}(c\rho) + \dfrac{13}{2}\sech^{2}(c\rho)\tanh(c\rho) - \dfrac{5}{2}\sech^{4}(c\rho)\right).
\label{DensidadeEnergia}
\end{equation}
The Fig. \ref{Fig-DensidadeEnergia} shows the energy density for some values of the evolution parameter $c$.
%Since $c$ is related to the cosmological constant, the Ricci flow governs how a flow of the Bulk cosmological constant changes the source of this smoothed string-like braneworld \cite{Charuto}. 
The source is shifted from the origin, and the width $\epsilon$ can be estimated as
$\epsilon = \bar{\rho}$, where $\bar{\rho}$ stands to the half-maximum position of $t_0$.
Further, as more the value of $c$ less the width of the core $\epsilon$ and then, closer to the GS model is the string-cigar solution.

%The energy density distribution introduces the thickness of the brane which extends from $\rho = 0$ to
%Note that the maximum of $t_0$ is displaced from the origin, allowing to define the core of the brane at this position. The region where $t_0$ is zero (far from the origin) may be though as vacuum.
%===============| Densidade de Energia |=======================================
%It's maximum is displaced from the origin and the thickness of the $3-$brane may be realised.
%=====================================================================

%The gravity localization is made by the same procedure as on GS model. The resulting radial equation for the graviton is
In the string-cigar geometry, the radial KK equation reads
\begin{equation}
\label{massivemodeequation2}
\phi_m''+c\left[-\frac{5}{2}\tanh^{2}{(c\rho)} +  \hspace{0.1cm} \frac{\text{sech}^{2}(c\rho)}{\tanh{(c\rho})}\right]\phi_m'+\e^{(c\rho-\tanh{(c\rho)})}\left(m_0^{2}-\frac{l^{2}c^{2}}{
\tanh^ { 2 } {(c\rho)}} \right)\phi_m = 0.
\end{equation}

Note that the mass term has a radial dependence which blows up at the origin and converges to
the GS value asymptotically. For $\rho \rightarrow \infty$, the equation (\ref{massivemodeequation2}) assumes the same
form as (\ref{Sturm-Liouville-GS}) with a re-scaled mass $m \rightarrow \e^{-1/2} m$. Hence,
the asymptotic solutions of (\ref{massivemodeequation2}) have the same behavior of Eq. (\ref{GS-MassiveModes}) \cite{Charuto}.

A normalized zero-mode solution satisfying the Schr\"{o}dinger equation for the
string-cigar model, is defined as \cite{Charuto}
\begin{equation}
\Psi_0(\rho,c) = N\sigma(\rho,c)\left(\dfrac{\tanh c\rho}{c}\right)^{1/4},
\label{Charuto-ZeroMode}
\end{equation}
where $N$ is a normalization factor. The solution (\ref{Charuto-ZeroMode}) can be realized as a
smoothed GS massless mode (\ref{GSzeromode}). In contrast with the GS model, the string-cigar
massless mode vanishes at the origin because of the conical behavior. This agrees with the
shift of the brane core.

However, the complexity of the radial equation (\ref{massivemodeequation2})
turns necessary the use of numerical methods to derive solutions for the complete domain. They are presented on section \ref{NumericalResults-1}.

%%%%%%%%%%%%%%%%%%%%%%%%%%%%%%%%%%%%%%%%%%%%%%%%%%%%%%%%%%%%%%%%%%%%%%%%%%%%%%%%%%%%%%%%%%%%%%%%%%%%%%%%%%%%
%%%%%%%%%%%%%%%%%%  NUMERICAL RESULTS (ESPECTRUM AND EIGENFUNCTIONS) %%%%%%%%%%%%%%%%%%%%%%%%%%%%%%%%%%%%%%%
%%%%%%%%%%%%%%%%%%%%%%%%%%%%%%%%%%%%%%%%%%%%%%%%%%%%%%%%%%%%%%%%%%%%%%%%%%%%%%%%%%%%%%%%%%%%%%%%%%%%%%%%%%%%

\section{Mass spectrum and KK eigenfunctions}
\label{NumericalResults-1}

We solved the equations (\ref{Sturm-Liouville-GS}) and (\ref{massivemodeequation2}) by the matrix method \cite{MatrixMethod} with second order truncation error in order to attain the complete KK spectrum. Since the angular number $l$ leads to a degenerate spectrum, $l=0$ solutions (which is referred to s-waves) were searched. Henceforward, we will label the graviton mass as $m$, instead of $m_0$.
%Since the mass terms contain the angular momentum contribution providing a degeneracy for the GS model and a position-dependent mass for the String-Cigar %model, we look for $l = 0$ solutions referring to four-dimensional gravitons.
%Due the $\e^{c\rho}$ term, variations on parameters $c$ led to similar results for proportional domain variations, so we present solutions for $c = 1.0$.
Further, these Sturm-Liouville problems are extremely sensible on
$c$ parameter variations due to the exponential term. Then, to prevent overflow errors, we fixed $c = 1.0$.
%The machine precision is extrapolated when $k > 3.0$,
For the GS model, the domain $\rho$ was discretized in $[0.0,\hspace{0.07cm}5.7]$ with $N = 500$ uniform subdivisions.
However, since Eq. (\ref{massivemodeequation2}) is singular at the origin,
and  the $1$st ($0$th)-order derivative coefficient is strongly active for small (large) $\rho$ values, the optimum domain for the string-cigar model is $[0.01, \hspace{0.07cm} 5.70]$.

The number of subdivisions of the interval $[a,\rho_{\mx}]$, $n$, is intrinsically related with the number of the zeroes of the Bessel function $\tilde{n}$ in this range. In fact, as more $n$ more zeroes exist in the interval. These zeroes split the interval in a partition. Therefore, the eigenvalues $m_{n}$ are labelled by the number of subdivisions. Henceforward, $n$ and $\tilde{n}$ are equivalent labels.

%where the function coefficients have the same order of magnitude preventing us loss of ODE information.

% we can not initiate the grid for the String-Cigar model on $a = 0$. Moreover, . Due to this fact,
%The differential equation, with the boundary conditions (\ref{Cond-Cont-GS}), turns to a generalized eigenvalue problem for a $N\times N$ tri-diagonal matrix $\mathbf{\hat{A}}$:
%\begin{equation}
%\mathbf{\hat{A}}\vec{\mathbf{y}} = m^2 \mathbf{\hat{B}} \vec{\mathbf{y}},
%\end{equation}
%where
%$$\mathbf{\hat{A}} = \left( \begin{array}{ccccc}
%-2 & 2 &  &  &  \\
%1-2hp(x_1) & -2 & 1+2hp(x_1) &  &  \\
% & \ddots & \ddots & \ddots &   \\
% &  & 1 - 2hp(x_{N-1}) & -2 & 1+2hp(x_{N-1}) \\
% &  &  & 2 & -2
%\end{array}  \right),$$
%$$\vec{\mathbf{y}} = \left(\begin{array}{c}
%y(x_0) \\
%y(x_1) \\
%\vdots \\
%y(x_{N-1}) \\
%y(x_{N})
%\end{array}\right), \hspace{1cm} \mathbf{\hat{B}} = \left( \begin{array}{cccc}
%-h^2q(x_0) &  &  &  \\
% & -h^2q(x_1) &  &  \\
% &  & \ddots &  \\
% &  &  & -h^2q(x_N)
%\end{array}   \right). $$

\begin{figure}[!htb] % Duas figuras lado a lado
       \begin{minipage}[b]{0.48 \linewidth}
           \includegraphics[width=\linewidth]{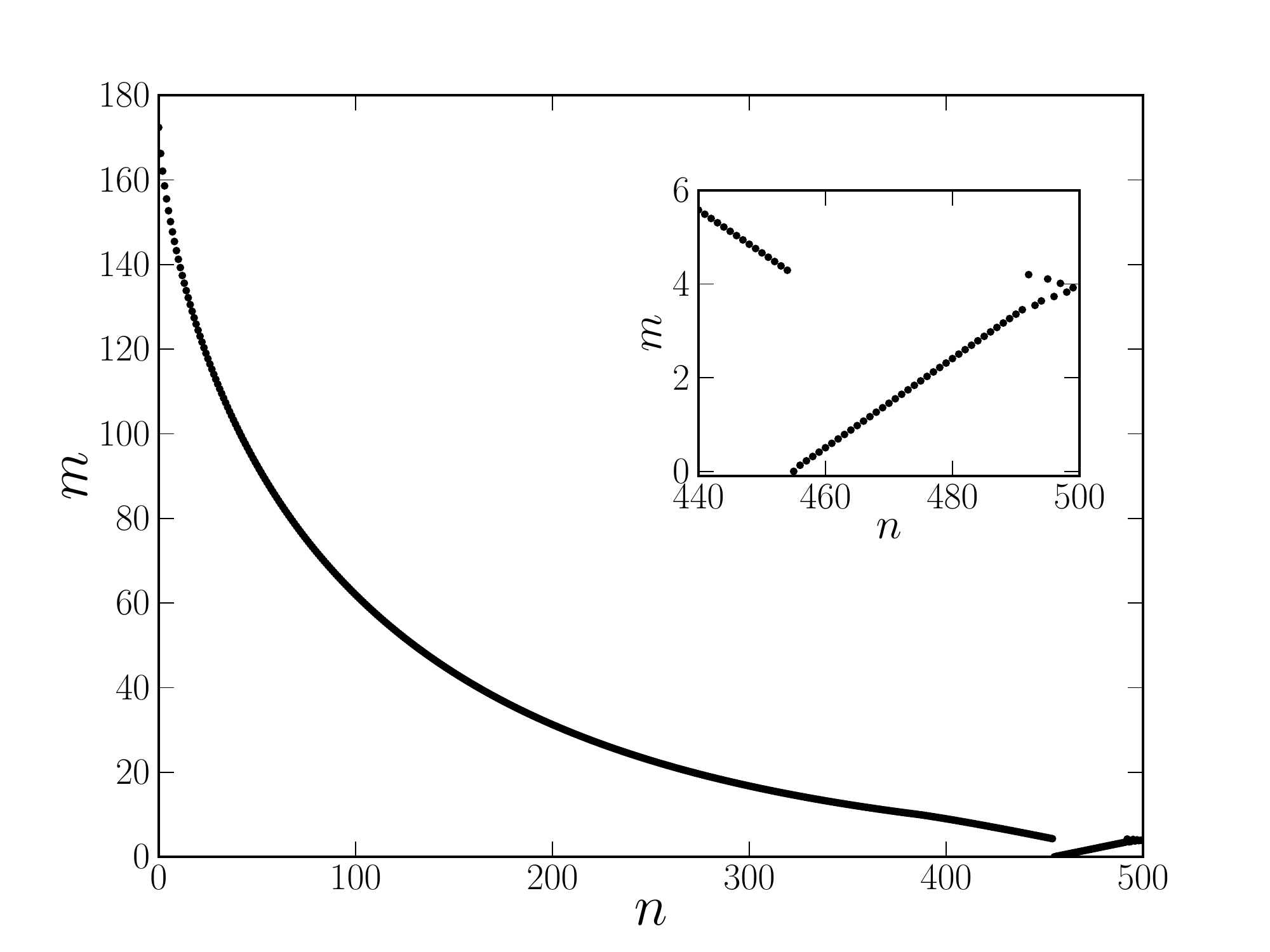}\\
           \caption{Thin string model's mass spectrum obtained by the matrix method. The sub-graph is the linear regime's scale magnification. The last points are numerical noises.}
          \label{Fig-Espectro-GS}
       \end{minipage}\hfill
       \begin{minipage}[b]{0.48 \linewidth}
           \includegraphics[width=\linewidth]{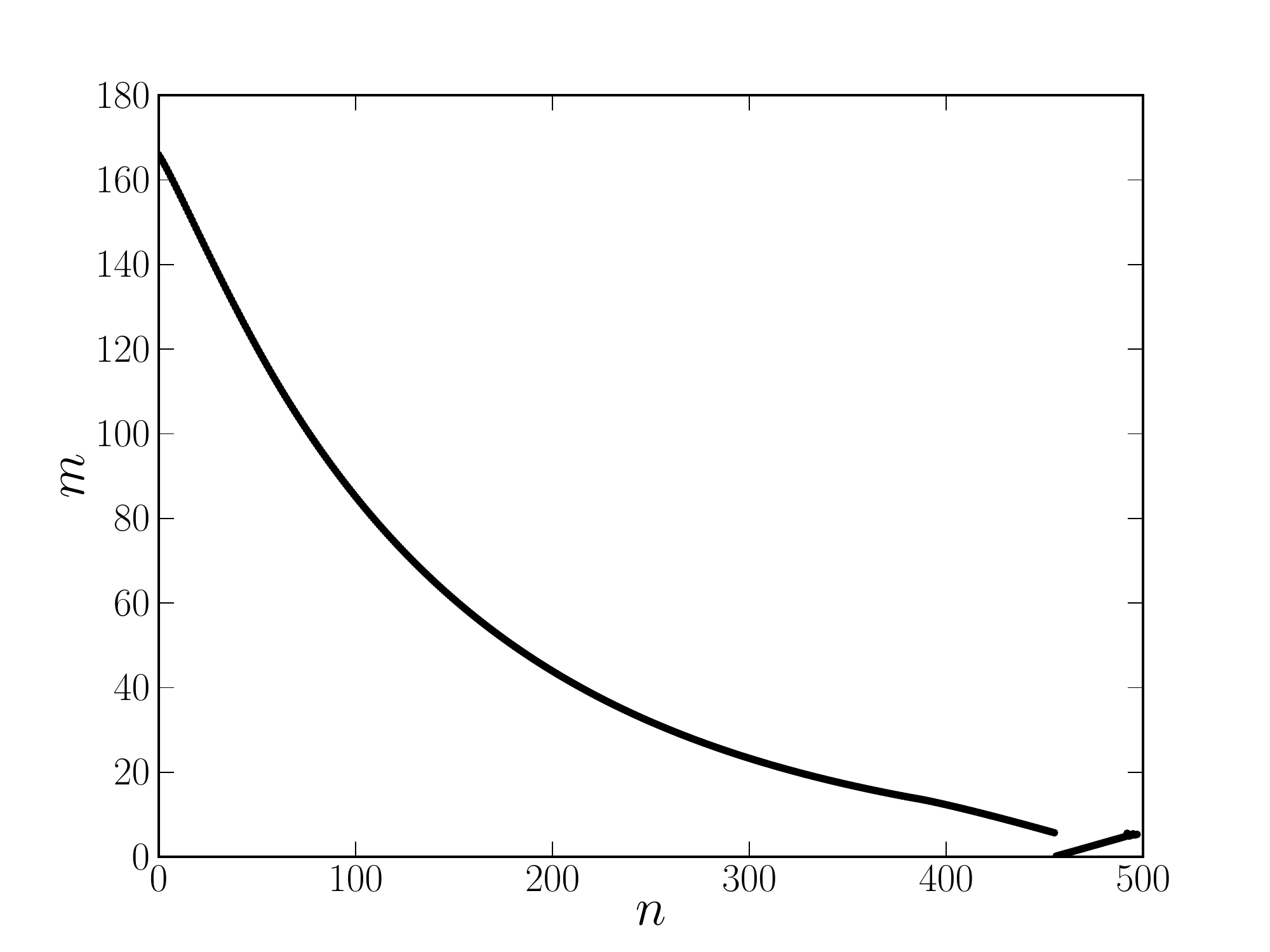}\\
           \caption{Mass spectrum of the string-cigar model obtained by matrix method. It is very similar to the GS model except by the absence of a transient state.}
           \label{Fig-Espectro-Charuto}
       \end{minipage}
   \end{figure}

The figures \ref{Fig-Espectro-GS} and \ref{Fig-Espectro-Charuto} show the complete
spectrum $m_n$ for GS and string-cigar models, respectively. The eigenvalues are all real and
thus, the models do not carry tachyons. For large values of the discretization index $n$
(on the present case, $n > 455$) the mass spectrum for both models grows linearly, what agrees
with the GS model \cite{GS}. On the other hand, for $n\leq 400$, the mass values decrease as
$m_{n}\approx \frac{1}{n}$, an important new behavior that can leads to corrections on the Newtonian potential \cite{GS}. The decreasing behavior for small $n$ matches with the analytical expression for the
zeros of the Bessel functions $J_{\frac{3}{2}}\left(\frac{2m_n}{c}\e^{\frac{1}{2}c\rho}\right)=0$, namely \cite{watson}
\begin{equation}
m_{n}=\frac{c}{2}\Big[\left(n-\frac{1}{2}\right)\pi +\frac{1}{4\Big[\left(n-\frac{1}{2}\right)\pi\Big]}+...\Big]\e^{-\frac{c\rho}{2}}.
\end{equation}
Hence, for small $n$, the decreasing behavior prevails over the linear regime.

\begin{figure}[h]
 \centering
    \includegraphics[width=0.65\textwidth]{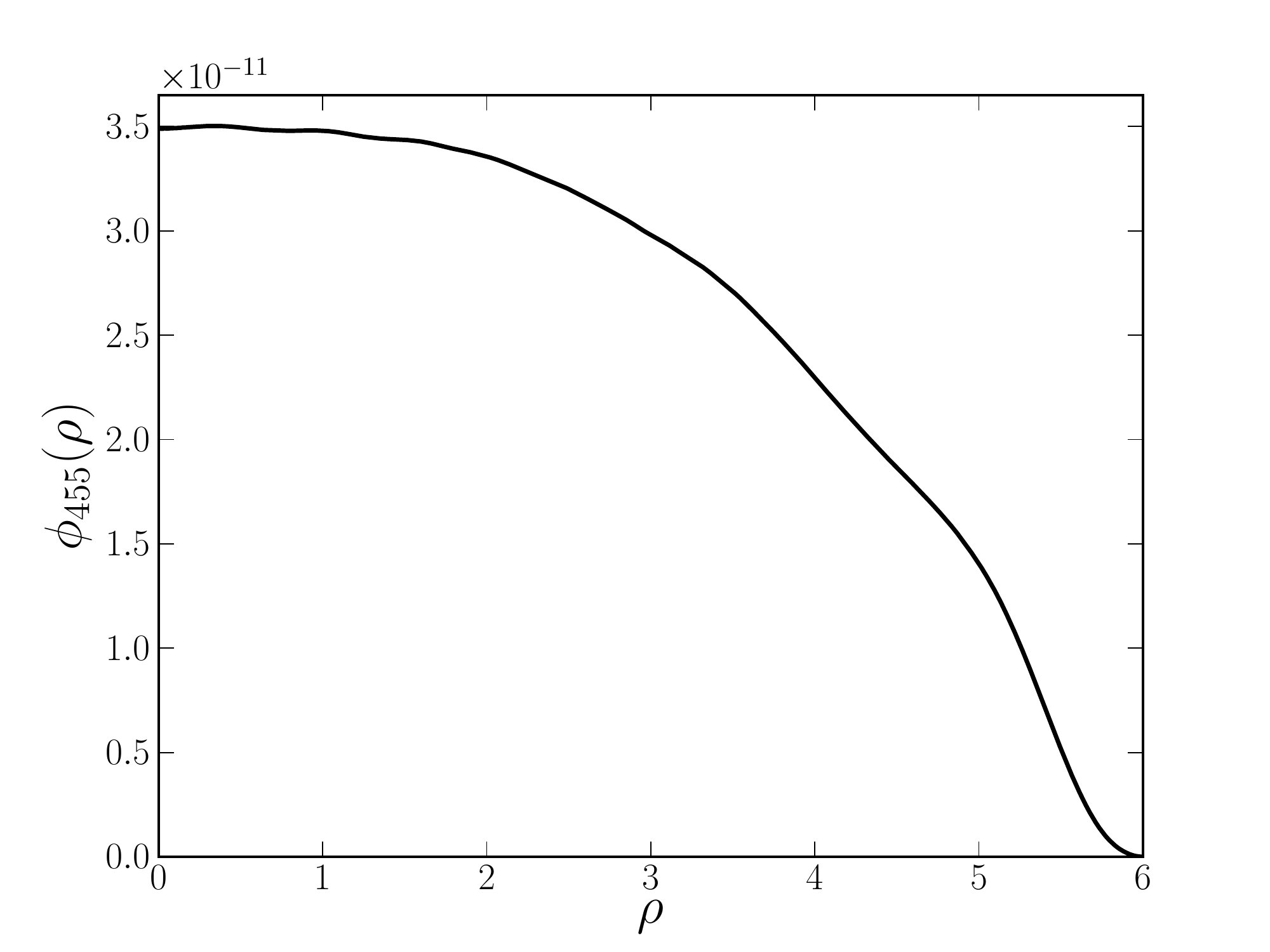}
 \caption{GS model's numerical eigenfunction for $m = 5.172\times 10^{-7}$.}
 \label{Fig-ModoZeroNumerico-GS}
\end{figure}

For $n = 455$, the minimum mass value is $m_{455} = 5.172\times 10^{-7}$, rather small
compared to the whole set, whose respective eigenfunction $\phi_{455}$ is plotted in Fig.
\ref{Fig-ModoZeroNumerico-GS}. This eigenfunction possesses a tiny amplitude and shares a
 similar profile of the GS localized massless mode. Such feature can be reported  as a transient
 state. However, unlike the GS zero mode, $\phi_{455}$ satisfies the boundary
conditions (\ref{Cond-Cont-GS}).

The numerical analysis revealed another feature of the GS model not present in the
 Ref. \cite{GS}, a mass gap between the decreasing and the linear regimes.
 It turns out that there are two asymmetric mass gaps on GS model between the transient state
 $m_{455}$ and its neighbours. Graphically, the gap between the decreasing regime and the transient state is clearly noticeable. The table (\ref{tabela}) shows that the separation between $m_{455}$ and $m_{456}$ is higher than
the subsequent ones that are approximately $\Delta m \approx 0.094$. Therefore, the transition between the decreasing and linear regimes is not continuous and passes through a transient state of tiny mass.
\begin{center}
\begin{table}
\begin{tabular}{|c|c|}
\hline
$m_{454} = 4.295$ 				 & $m_{458} = 0.317$  \\
\hline
$m_{455} = 5.172 \times 10^{-7}$ &  $m_{459} = 0.411$ \\
\hline
$m_{456} = 0.130$ 				 & $m_{460} = 0.504$  \\
\hline
$m_{457} = 0.224$ 				 & $m_{461} = 0.598$ \\
\hline
\end{tabular}
\caption{Some mass eigenvalues for high values of the discretization index where the linear regime (for $n > 455$) is noticiable. The smallest mass value $m_{455}$ corresponds to the transient state.}
\label{tabela}
\end{table}
\end{center}

The existence of a mass gap and the transient massive mode agrees with the analytical solutions for the massive modes (\ref{GS-MassiveModes}) and for
the massless mode (\ref{GSzeromode}). Indeed, for small masses it can expands the massive mode as
\begin{equation}
\label{GS-smallmasses}
\phi_{m\rightarrow 0}(\rho) = D_1\left(\frac{c}{2m}\right)^{\frac{5}{2}} + D_2 \left(\frac{c}{2m}\right)^{\frac{1}{2}}\e^{c\rho} + D_3\left(\frac{2m}{c}\right)^{\frac{1}{2}}\e^{\frac{3}{2}c\rho} + \mathcal{O}(\e^{\frac{5}{2}c\rho}),
\end{equation}
where $D_1$, $D_2$ and $D_3$ are constants determined by the constants $B_1$ and $B_{2}$ and by the expansion of the Bessel functions at the origin. It is worthwhile to say that
since $\lim_{m\rightarrow 0}\phi_{m}(\rho)=+ \infty$, it is not possible to attain continuously the massless mode (\ref{GSzeromode}) from the massive mode (\ref{GS-MassiveModes}). Thus, there is a mass gap between the massless and the first massive modes. Further, the small amplitude of $\phi_{455}$ can be explained by the small values of the constants, e.g., $D_3\approx 10^{-32}$, and by the boundary conditions which relate the constants.

The string-cigar mass spectrum is very similar to the GS one except by the absence of the tiny transient massive mode as in GS model. The minimum mass value is far from zero. Thus, the near brane resolution driven by the string-cigar geometry avoids this transient mode, and it becomes more regular the massive spectrum. Furthermore, it turns out that the mass gap remains invariant for all the values of $c$ used. Then, this suggests that the present geometrical flow do not break  the discontinuity between the two regimes.

%===============================| Espectra |=======================================

%\begin{figure}[htb] % Duas figuras lado a lado
%        \begin{minipage}[b]{0.4 \linewidth}
%            \fbox{\includegraphics[width= \linewidth]{Espectro-GS.pdf}}\\
%            \caption{Thin string model's mass spectrum obtained by the matrix method. The sub-graph is the linear regime's scale magnification. The last points are numerical noises.}
%            \label{Fig-Espectro-GS}
%        \end{minipage}\hfill
%        \begin{minipage}[b]{0.4 \linewidth}
 %           \fbox{\includegraphics[width= \linewidth]{Espectro-Charuto.pdf}}\\
%            \caption{Mass spectrum of the String-Cigar model obtained by matrix method. It is very similar to the GS model except by the absence of a transient state.}
%            \label{Fig-Espectro-Charuto}
%        \end{minipage}
%\end{figure}

%==========================================================================================

%===========================| Modo-Zero Numerico GS|=======================================
%%=========================================================================================

% An evidence for a localized zero-mode  satisfying the Neumann boundary conditions.
%===========================| GS eigenfunctions |==========================================

%\begin{figure}[htb] % Duas figuras lado a lado
%        \begin{minipage}[b]{0.48 \linewidth}
%            \fbox{\includegraphics[width=\linewidth]{Autofuncao-GS-01.eps}}\\
 %           \caption{Thin string GS model's numerical eigenfunction for $m = 37.156$.}
%           \label{Fig-GS-Eigenfunction-01}
%        \end{minipage}\hfill
%        \begin{minipage}[b]{0.48 \linewidth}
%%            \fbox{\includegraphics[width=\linewidth]{Autofuncao-GS-02.eps}}\\
 %           \caption{Thin string GS model's numerical eigenfunction for $m = 22.729$.}
 %           \label{Fig-GS-Eigenfunction-02}
%        \end{minipage}
%\end{figure}

\begin{figure}[!htb] % Duas figuras lado a lado
       \begin{minipage}[b]{0.48 \linewidth}
           \includegraphics[width=\linewidth]{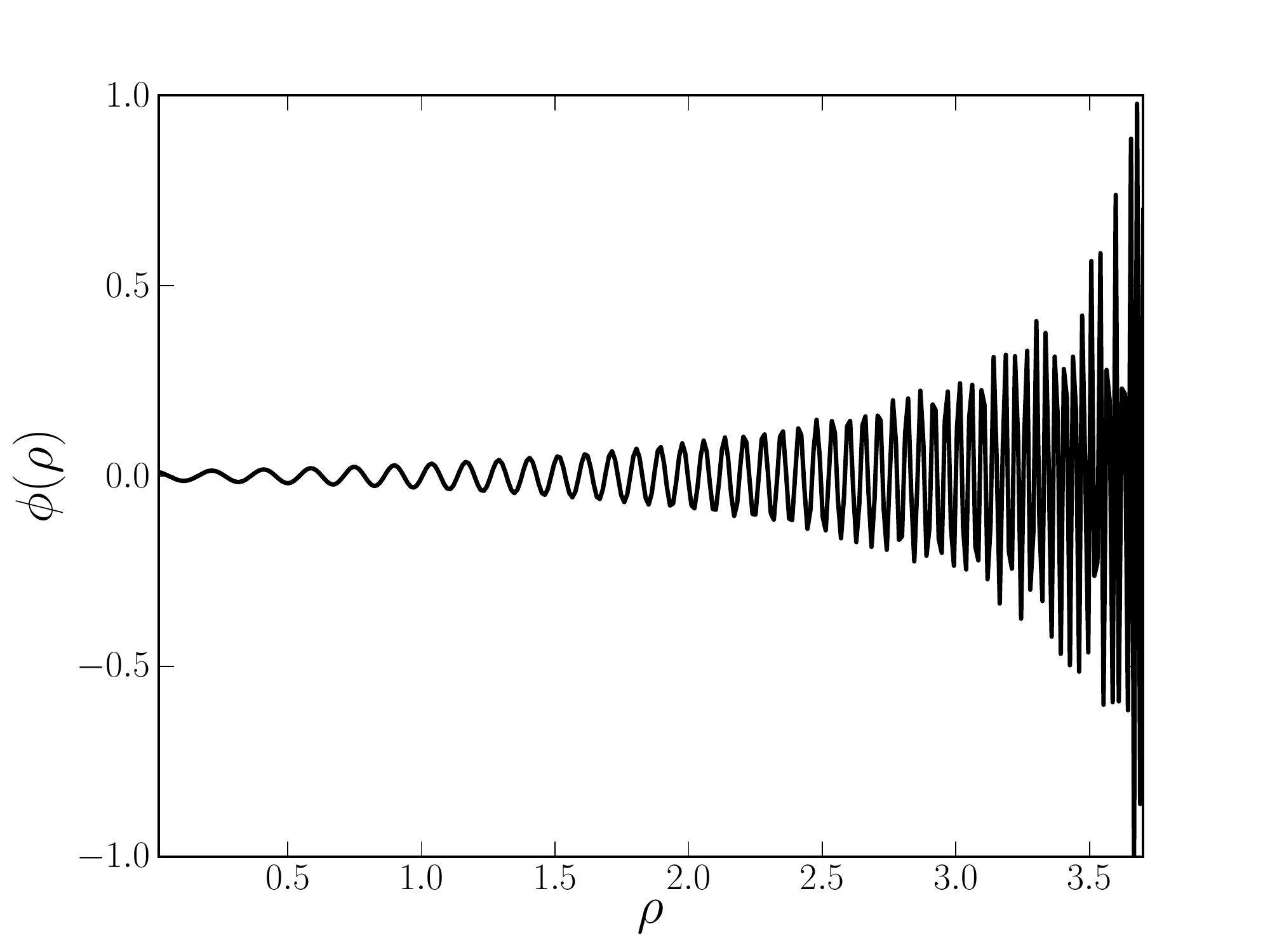}\\
           \caption{Thin string GS model's numerical eigenfunction for $m = 37.156$.}
          \label{Fig-GS-Eigenfunction-01}
       \end{minipage}\hfill
       \begin{minipage}[b]{0.48 \linewidth}
           \includegraphics[width=\linewidth]{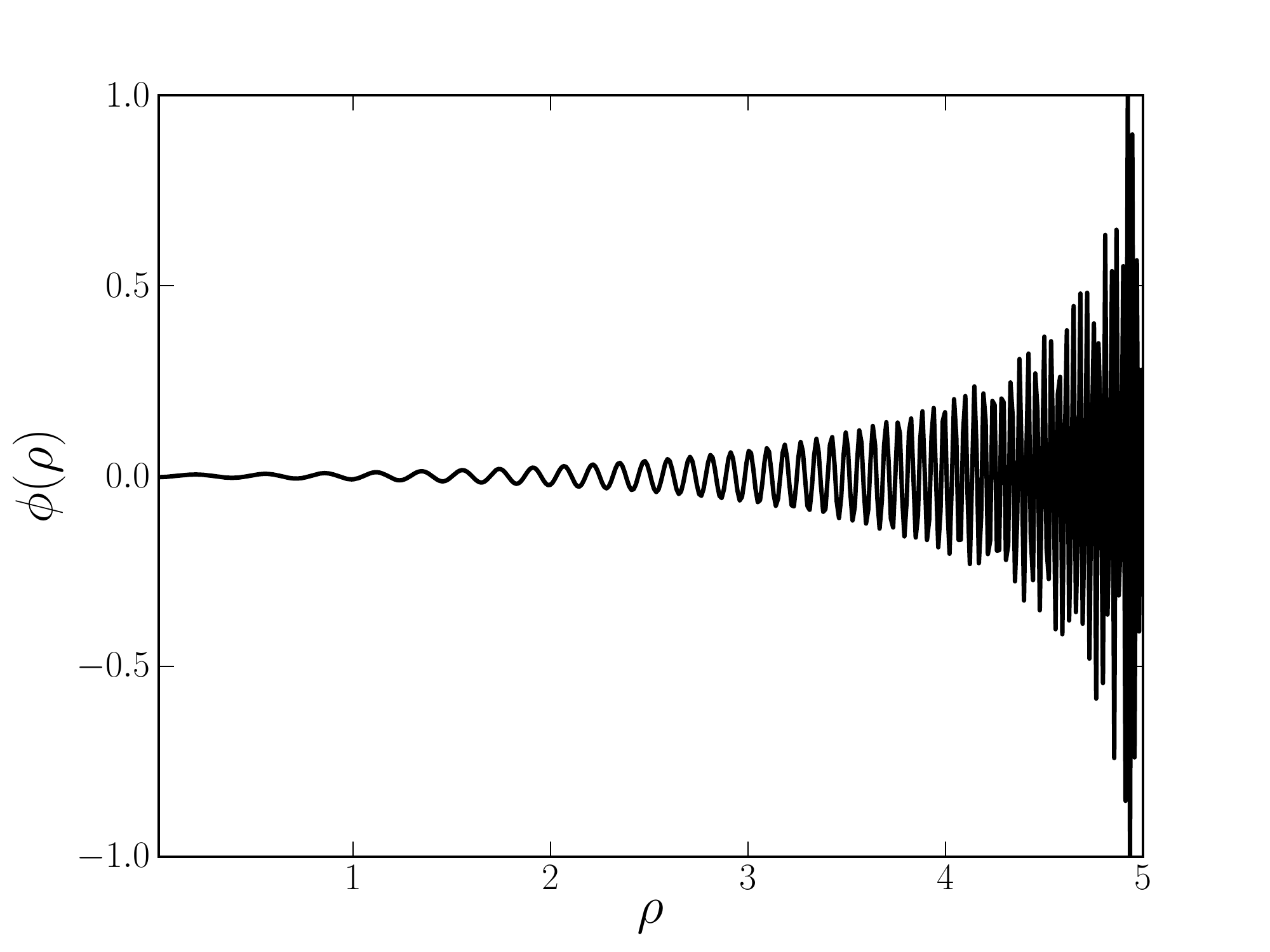}\\
           \caption{Thin string GS model's numerical eigenfunction for $m = 22.729$.}
           \label{Fig-GS-Eigenfunction-02}
       \end{minipage}
   \end{figure}

%==========================================================================================

\begin{figure}[!htb] % Duas figuras lado a lado
       \begin{minipage}[b]{0.48 \linewidth}
           \includegraphics[width=\linewidth]{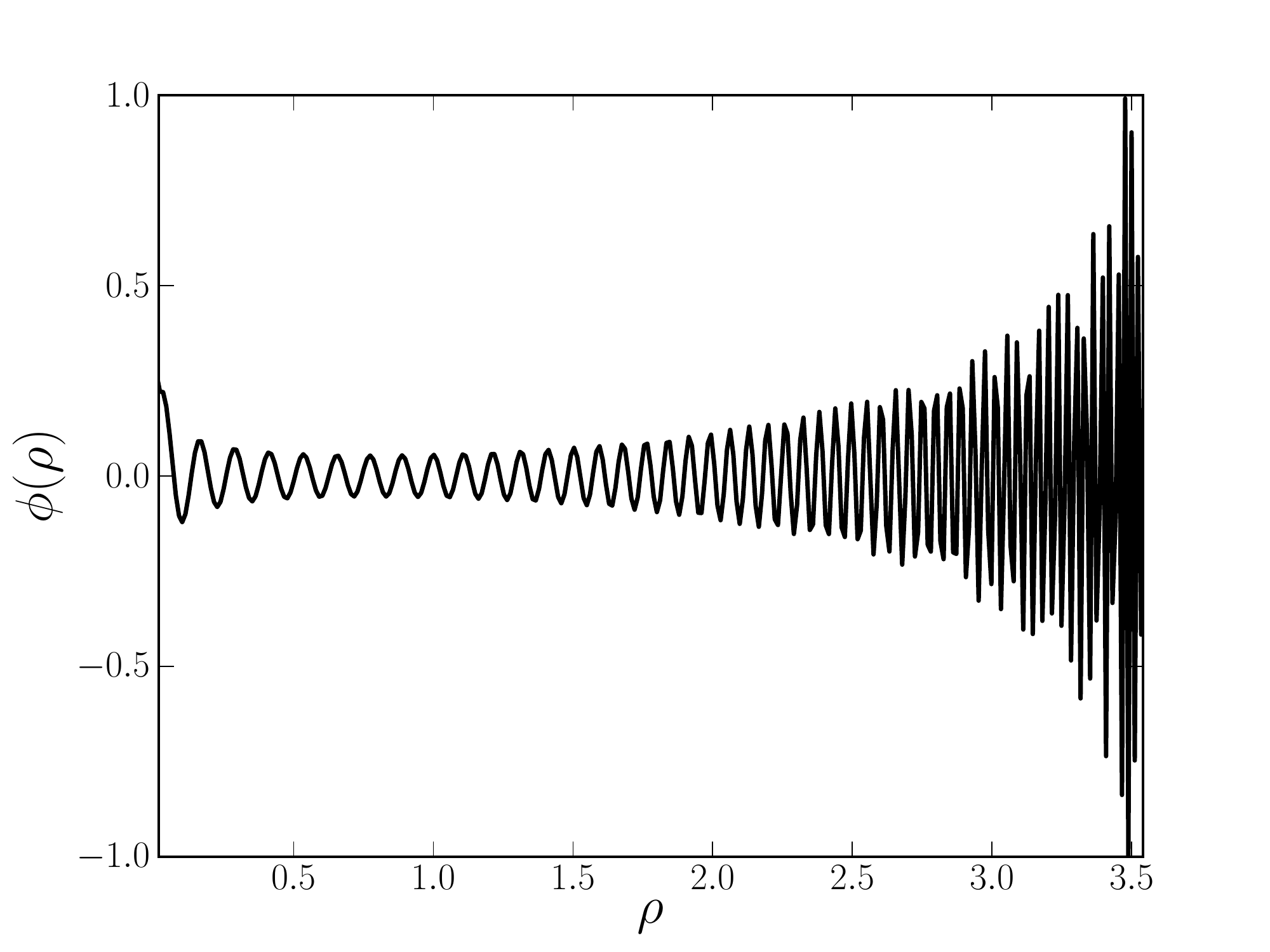}\\
           \caption{String-cigar model's numerical eigenfunction for $m = 37.022$.}
          \label{Fig-Charuto-Eigenfunction-01}
       \end{minipage}\hfill
       \begin{minipage}[b]{0.48 \linewidth}
           \includegraphics[width=\linewidth]{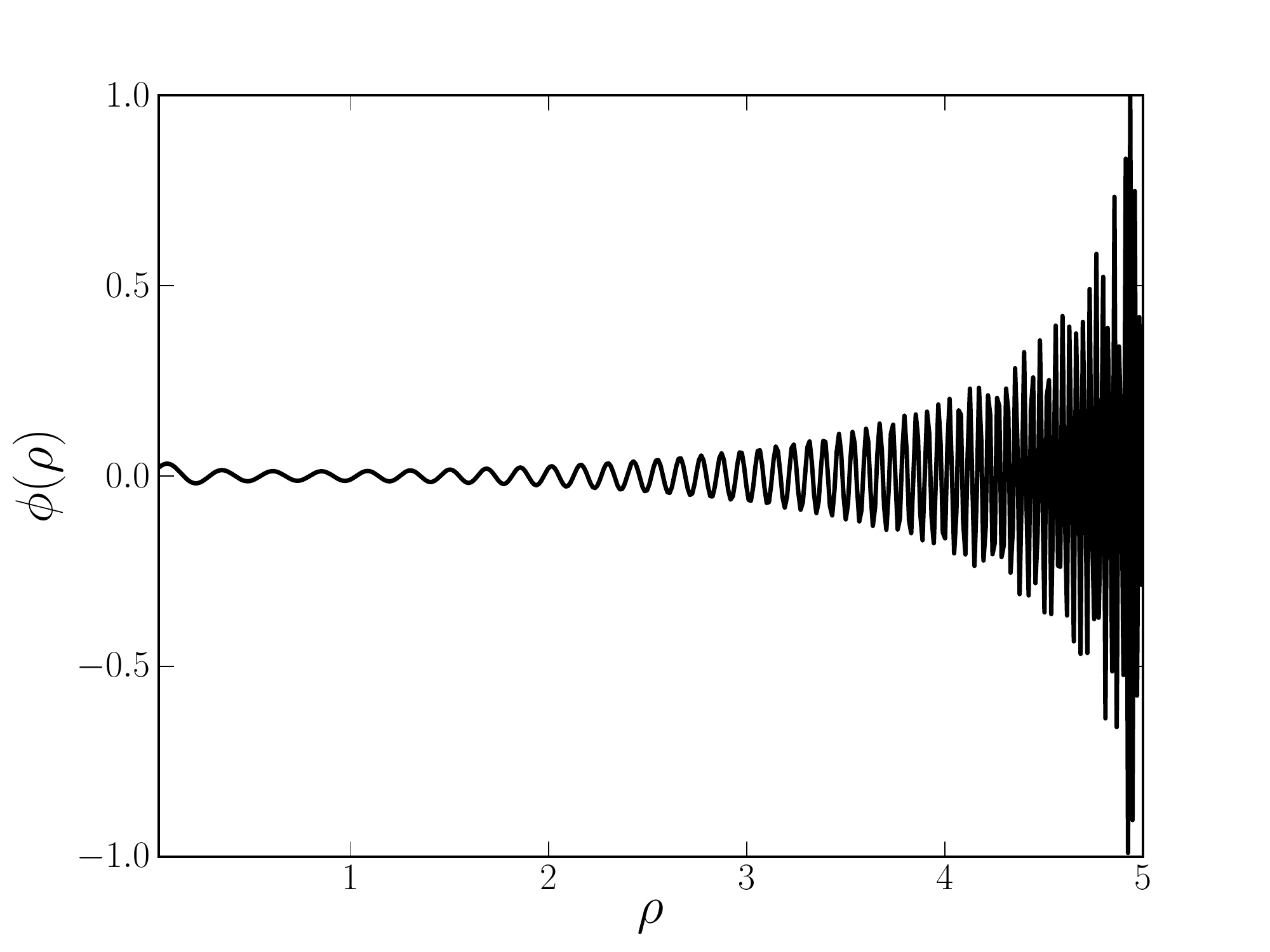}\\
           \caption{String-cigar model's numerical eigenfunction for $m = 22.721$.}
           \label{Fig-Charuto-Eigenfunction-02}
       \end{minipage}
   \end{figure}

The GS and the string-cigar eigenfunctions were plotted for some mass eigenvalues in the Figs. \ref{Fig-GS-Eigenfunction-01}, \ref{Fig-GS-Eigenfunction-02}, \ref{Fig-Charuto-Eigenfunction-01} and \ref{Fig-Charuto-Eigenfunction-02}, respectively.
%For the GS model, all solutions for $n \neq 455$  have the behavior of the Bessel functions,
%as shown in .
% are plotted on figures .
As expected, the massive solutions for both models match the same behavior asymptotically.
Near the origin, the massive modes of the string-cigar model have a bigger amplitude compared with those of the GS model. Moreover,
they are smooth in the core and near the brane as predicted in Ref. \cite{Charuto}.
At this limit, the string-cigar eigenfunctions behave as Bessel functions of first kind $J_{0}(m\rho)$. Indeed, a first-order expansion of the coefficients of Eq. (\ref{massivemodeequation2}) at the origin yields to

\begin{equation}
\label{cigarnearbrane}
\phi''_{m}+\left(\frac{1}{\rho}-\frac{2}{3}c^{2}\rho\right)\phi'_{m}+m^{2}\phi_{m}=0.
\end{equation}
For $\rho\approx 0$ the term $\frac{1}{\rho}$ prevails over the term $-\frac{2}{3}c^{2}\rho$ what yields to a Bessel equation
whose solution is
\begin{equation}
\phi_{\rho\rightarrow 0}(\rho) = E_{1}J_{0}(m\rho) + E_{2}Y_{0}(m\rho),
\end{equation}
where $E_1$ and $E_2$ are constants. Since $Y_{0}$ diverges at the origin, $E_{2} = 0$,  fitting the behavior sketched in Figs. \ref{Fig-Charuto-Eigenfunction-01} and \ref{Fig-Charuto-Eigenfunction-02}.

%===============| String-Cigar eigenfunctions |===================================

%\begin{figure}[htb] % Duas figuras lado a lado
%       \begin{minipage}[b]{0.48 \linewidth}
%            \fbox{\includegraphics[width=\linewidth]{Autofuncao-Charuto-01.eps}}\\
%            \caption{String-Cigar model's numerical eigenfunction for $m = 37.022$.}
%            \label{Fig-Charuto-Eigenfunction-01}
%        \end{minipage}\hfill
%        \begin{minipage}[b]{0.48 \linewidth}
%            \fbox{\includegraphics[width=\linewidth]{Autofuncao-Charuto-02.eps}}\\
%            \caption{String-Cigar model's numerical eigenfunction for $m = 22.721$.}
%            \label{Fig-Charuto-Eigenfunction-02}
%        \end{minipage}
%\end{figure}

%==========================================================================================

%%%%%%%%%%%%%%%%%%%%%%%%%%%%%%%%%%%%%%%%%%%%%%%%%%%%%%%%%%%%%%%%%%%%%%%%%%%%%%%%%%%%%%%%%%%%%%%%%%%%%%%%%%%%
%%%%%%%%%%%%%%%%%%%%%%%%%%%  NUMERICAL RESULTS RESONANT MODES%%%%%%%%%%%%%%%%%%%%%%%%%%%%%%%%%%%%%%%%%%%%%%%
%%%%%%%%%%%%%%%%%%%%%%%%%%%%%%%%%%%%%%%%%%%%%%%%%%%%%%%%%%%%%%%%%%%%%%%%%%%%%%%%%%%%%%%%%%%%%%%%%%%%%%%%%%%%

\section{KK resonant modes}
\label{NumericalResults-2}

%It is useful for the study of gravitational massive states the Schroedinger formalism. It provides a quantum analogue potential
%\cite{CohenKaplan-1999}.
%This idea was introduced by ref.\cite{Rubakov-Shaposhnikov} and w

Despite the divergent behavior of the massive gravitational modes, some of these modes can show a resonant profile \cite{Gremm, GravityBloch}. These states may be found by means of the resonance method \cite{CASA-Fermion-TwoField-ThickBrane} that provides solutions of the Schr\"{o}dinger equation which exhibit large amplitudes near the brane in comparison with its values far from the defect. Large peaks in the distribution of the wave function in terms of the energy reveal the existence of resonant states \cite{CASA-Fermion-TwoField-ThickBrane}.

%This phenomenon is only possible for $m^2 < U_{\mx}$, where $U_{\mx}$ is the maximum of the potential. In fact, solutions of the Schroedinger equation (\ref{EqSchroedinger}) for $m^2 \gg U_{\mx}$ are plane waves.

The method consists on defining the probability $P(m)$ to find a particle with mass $m$ at the position $z_0$ as
\begin{equation}
P(m) = \dfrac{|\Psi_m(z_0)|^2}{ \int_{z_{\mn}}^{z_{\mx}} |\Psi_m(z)|^2dz},
\label{Probabilidade}
\end{equation}
where $z_{\mn}$ and $z_{\mx}$ stands to domain limits. An extension of this idea was proposed on Ref. \cite{Ress-Chineses-Fermions-deSitter}, where a relative probability is defined as
\begin{equation}
P(m) = \dfrac{ \int_{z_a}^{z_b} |\Psi_m(z)|^2 dz}{ \int_{z_{\mn}}^{z_{\mx}} |\Psi_m(z)|^2 dz},
\label{ProbabilidadeRelativa}
\end{equation}
evaluated on a narrow range $[z_a,z_b]$.

\begin{figure}[!htb] % Duas figuras lado a lado
       \begin{minipage}[b]{0.48 \linewidth}
           \includegraphics[width=\linewidth]{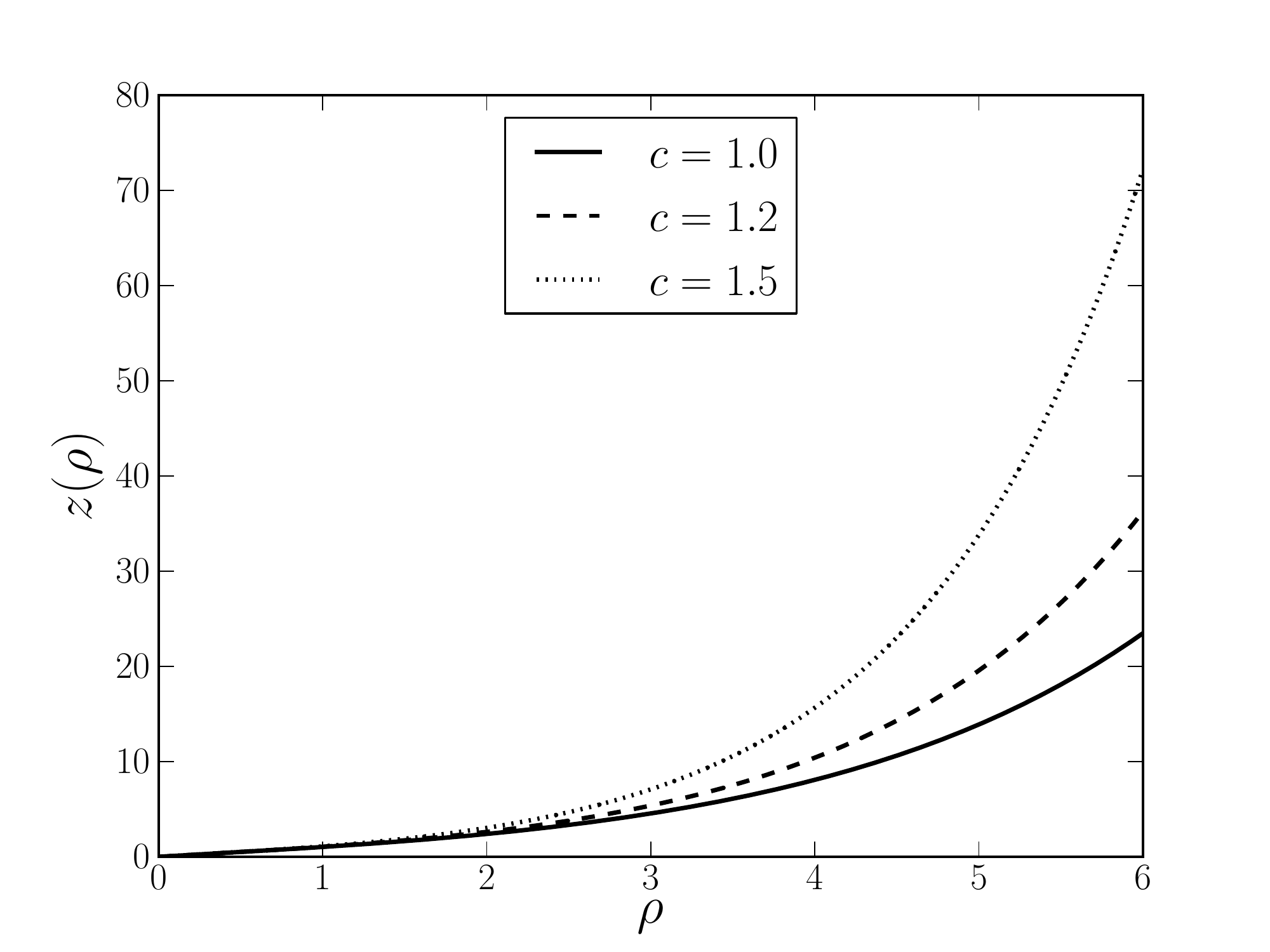}\\
           \caption{Numerical integral solution of the transformation $z(\rho)$.}
          \label{Fig-Transformacao-Z}
       \end{minipage}\hfill
       \begin{minipage}[b]{0.48 \linewidth}
           \includegraphics[width=\linewidth]{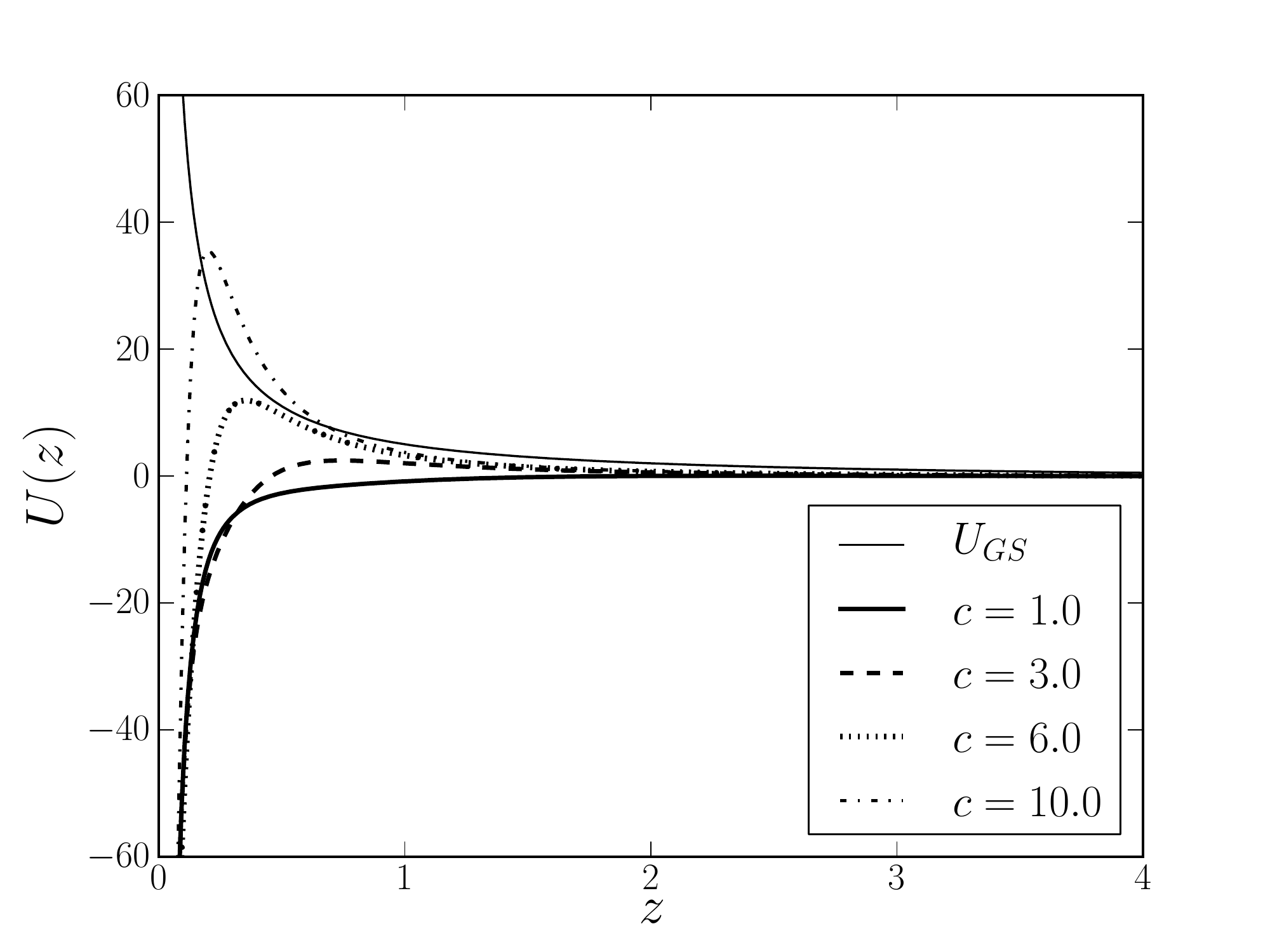}\\
           \caption{The analogue quantum potential $U(z)$.}
           \label{Fig-Potencial}
       \end{minipage}
   \end{figure}

Probability interpretations would be possible for the Sturm-Liouville eigenfunctions by defining the inner product with the weight function included.
However, the change of variable $z=z(\rho)$ acts as a domain stretch that improves the treatment for large $c$. By numerical integration, the change of coordinate is plotted in Fig. \ref{Fig-Transformacao-Z}. The analogue quantum potential for the GS model is
\begin{equation}
U_{GS}(z) = 6 z^{-2}.
\end{equation}
This potential function does not support bound state and then, there is not any probability to find massive gravitons on the infinitely thin string-like braneworld. Nevertheless, for the string-cigar model, the Schr\"{o}dinger potential, plotted in Fig. \ref{Fig-Potencial}, presents an infinite potential well what suggests the possibility of resonant modes.
% the integral on Eq.(\ref{Z-Transformation}) is difficult to solve for the String-Cigar warp factors (\ref{Charuto-WarpFactor}) and (\ref{Charuto-AngularFactor}). On reference \cite{Charuto}, the potential is sent back to the $\rho$ coordinate, which plot shows a deep well near origin allowing probability interpretation to find a massive graviton state on the brane. As the Schroedinger equation (\ref{EqSchroedinger}) must to be solved on the $z$ coordinate, we resort numerical methods to study these resonant states.
%As mentioned before, the numerical calculations are impossible for high values of the  cosmological constant. However, the Schroedinger reduced equation (\ref{EqSchroedinger}) admits a numerical treatment for higher values of the parameter $c$. It depends only on the numerical integral (\ref{Z-Transformation}) which solution, for some values of $c$, is presented on  This transformation . After this numerical integration, we may construct the potential function $U(z)$ plotted on fig.(\ref{Fig-Potencial}) for different values of $c$. The potential function has a negative infinite deep well suggesting the presence of tachyons. However, our numerical methods have not exhibited any negative value for
%$m^2$. In fact, our grid must not initiate on $a = 0$, so $U(a)\neq -\infty$.
For $c = 1.0$ the potential for the string-cigar model behaves as a Coulomb-like potential and, when the parameter $c$ increases, a barrier arises and the potential assumes the usual volcano shape.
%The maximum of the potential barrier is in accordance with the displaced maximum of the stress-energy tensor which indicates the position of the core of the brane.
Massive particles confined on the well may be interpreted as gravitational resonant states (massive quasi-localized gravitons highly coupled to the brane) \cite{Gremm, Csaki:2000fc, Gravity-Quasi-Ress}.

%===============| Fig - Potencial |===================================

%\begin{figure}[htb] % Duas figuras lado a lado
%        \begin{minipage}[b]{0.48 \linewidth}
%            \fbox{\includegraphics[width=\linewidth]{Z_Numerico.pdf}}\\
%            \caption{Numerical Integral solution of the transformation $z(\rho)$.}
%            \label{Fig-Transformacao-Z}
%        \end{minipage}\hfill
%        \begin{minipage}[b]{0.48 \linewidth}
%            \fbox{\includegraphics[width=\linewidth]{Potencial.pdf}}\\
%            \caption{The analogue quantum potential $U(z)$.}
%            \label{Fig-Potencial}
%        \end{minipage}
%\end{figure}

%==========================================================================================

We solved the Schr\"{o}dinger-like equation (\ref{EqSchroedinger}) using the Numerov method \cite{Numerov1, Numerov2}.
%Furthermore, the search for solutions with the highest probabilities  to find massive states near the brane is optimized by the resonance method \cite{Gravity-Quasi-Ress, CASA-Fermion-TwoField-ThickBrane, Wilami-FermionDeformedThickBrane, Makarius}.
The relative probability function (\ref{ProbabilidadeRelativa}) is more suitable to detect very narrow resonances if they exist \cite{Wilami-Gravitons-ThickBrane}. According to the energy-density given in Eq. (\ref{DensidadeEnergia}) the brane distribution is $[z_a, z_b] = [0.01, 0.50]$.
% which maximum position is $\bar{\rho} = 0.50$ for $c = 1.0$.
The domain is chosen to be $[z_{\mn},z_{\mx}] = [0.01, 5.0]$ ($10$ times the integration range), for which the plane wave probability would be $P(m) = 0.1$ \cite{Ress-Chineses-Fermions-deSitter}. The position of the resonance peak, where the physical information is stored, does not depend on $z_{\mx}$, since it is chosen sufficiently large \cite{Wilami-Gravitons-ThickBrane}.

\begin{figure}[htb] % Duas figuras lado a lado
        \begin{minipage}[b]{0.48 \linewidth}
            \includegraphics[width=\linewidth]{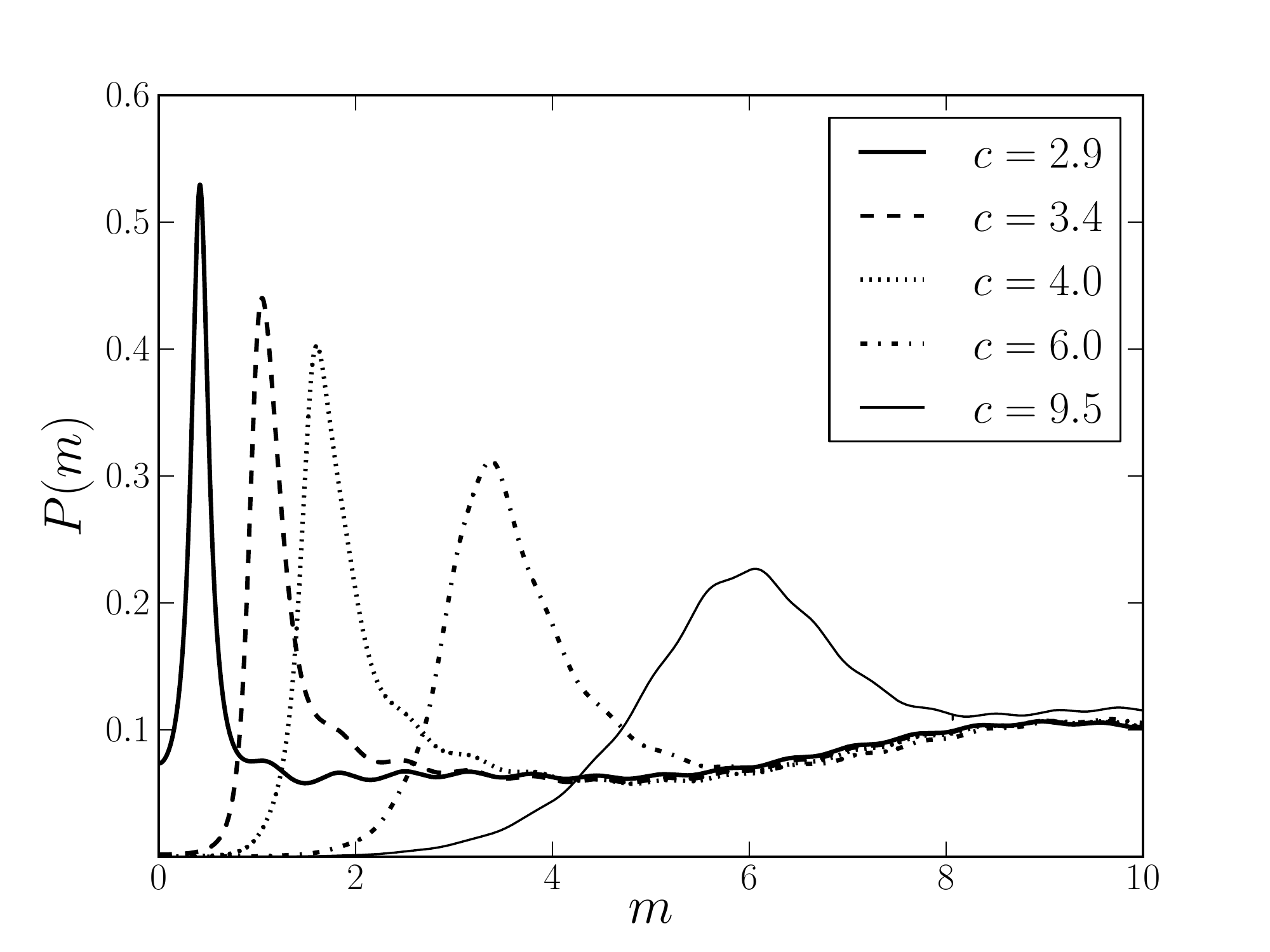}\\
            \caption{Plot of the relative probability $P(m)$. The peaks reveal the resonant massive states. For $m^2 \gg U_{\mx}$, a plateau is formed at $P = 0.1$ corresponding to plane wave regime.}
            \label{Fig-Ressonancia}
        \end{minipage}\hfill
        \begin{minipage}[b]{0.48 \linewidth}
            \includegraphics[width=\linewidth]{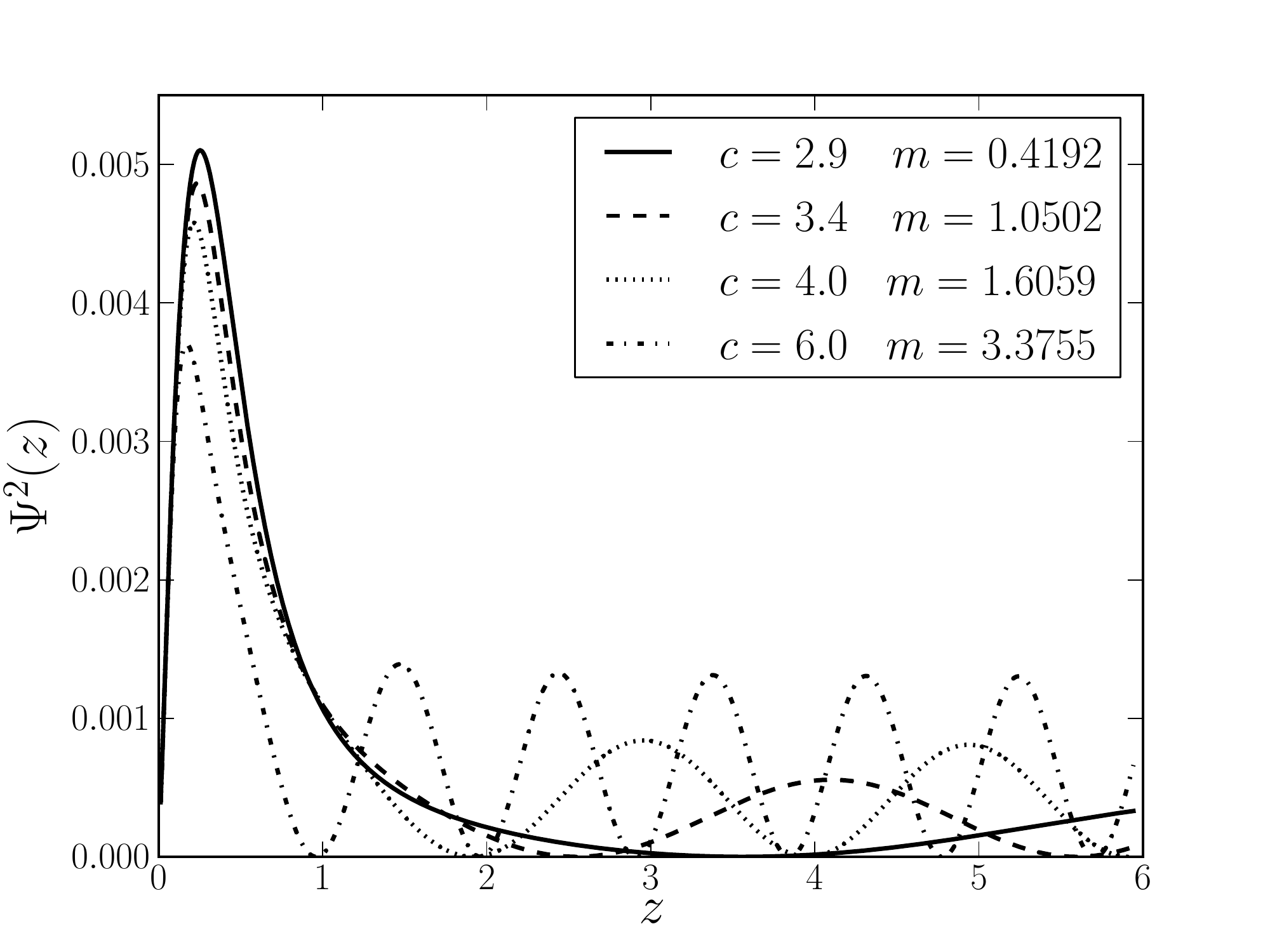}\\
            \caption{Solutions of the Schr\"{o}dinger-like equation for the masses corresponding to resonant peaks. The first solution has the highest probability to interact with the brane.}
            \label{Fig-Funcao-Onda}
        \end{minipage}
\end{figure}

The Fig.  \ref{Fig-Ressonancia} presents the numerical solution of the relative probability $P(m)$. The first (and largest) peak is revealed  for $c = 2.9$. The resonant peaks (there is only one for each $c$) decrease and become broader with $c$. This means that the resonant massive states lifetime (estimated as $(\Delta m)^{-1}$, where $\Delta m$ is the width at half maximum of the peak \cite{CASA-Fermion-TwoField-ThickBrane}) decreases with the cosmological constant.
%Indeed, as higher the bulk cosmological constant thinner the brane and then, closer to the GS model are the massive modes, where there is no resonant modes.

The Fig.  \ref{Fig-Funcao-Onda} exhibits the solutions of the Schr\"{o}dinger equation for the masses indicated with the probability peaks. The first solution shows that this particular massive graviton has the highest probability to be found on the brane. This solution shares a similar behavior to the zero-mode presented in Ref. \cite{Charuto} near the origin, albeit it oscillates asymptotically with large wavelength.

%From this result, we may argue that a KK gravitational mode may exist in our world as a resonant state. The reason why massive gravitons are not observed in
%our universe is that its coupling constant with the brane matter fields is very weak \cite{MassiveGravitons}.

%===============| Ressonancias |===================================

%==========================================================================================

The effects of the brane internal structure on the resonant modes are also shown in Fig.  \ref{Fig-Funcao-Onda}. Increasing the value of $c$, which makes the brane thinner (see Fig. \ref{Fig-DensidadeEnergia}), enhance the width of the resonance. Therefore, as $\epsilon\rightarrow 0$, the resonant states acquire shorter lifetimes reducing the possibility to massive gravitons result in any physical measurement. This result agrees with the fact that for $c\rightarrow \infty$ the potential barrier tends to the $U_{GS}$, where no massive state is allowed.

 %In fact, since the thin-string braneworld is a  limit of the String-Cigar one,

%At zero energy ($m=0$), the presence of a resonance is related to the existence of a large-distance region on which the $4$D laws of gravity are valid and if the resonance width becomes very large, it results in non-physical effects \cite{Gravity-Quasi-Ress}. Furthermore, the wavefunction for the nonzero modes calculated at the origin may contribute with a correction to the Newton's law on the $3-$brane \cite{GS}.

%%%%%%%%%%%%%%%%%%%%%%%%%%%%%%%%%%%%%%%%%%%%%%%%%%%%%%%%%%%%%%%%%%%%%%%%%%%%%%%%%%%%%%%%%%%%%%%%%%%%%%%%%%%%
%%%%%%%%%%%%%%%%%%%%%%%%%%%  CONCLUSIONS AND PERSPECTIVES %%%%%%%%%%%%%%%%%%%%%%%%%%%%%%%%%%%%%%%%%%%%%%%%%%
%%%%%%%%%%%%%%%%%%%%%%%%%%%%%%%%%%%%%%%%%%%%%%%%%%%%%%%%%%%%%%%%%%%%%%%%%%%%%%%%%%%%%%%%%%%%%%%%%%%%%%%%%%%%

\section{Conclusions and Perspectives}
\label{Conclusions}

This work displays a study of the gravitational Kaluza-Klein (KK) modes in the Gherghetta-Shaposhnikov (GS) and the string-cigar braneworlds.
By the analysis of the energy density, the former can be realised as a thin string-like model whereas the last is regarded as interior and exterior
smoothed GS model.

We obtain a new massless mode for the GS model by analysing the radial equation in the Schr\"{o}dinger approach. It turns out that this massless mode has a
bigger amplitude and rate of decay compared with that of the GS model. Further, the transition between the massive and massless modes is not smooth what yields to a mass gap.

We obtained via numerical analysis a complete spectrum of masses and eigenfunctions for both models. Besides the mass gap between the massless mode and the massive modes, both models present a gap separating two regimes. For large values of the discrete index $n$, the spectrum behaves linearly, as predicted by GS model. However, for small $n$, the masses decrease as $1/m_n$. This result can be explained by the feature of the zeros of the Bessel function of the first kind in the two regimes.

Asymptotically, the string-cigar eigenfunctions are similar to GS ones whereas near the brane they behave as the Bessel function of the first kind. It turns out that the amplitude of the modes is bigger in the string-cigar model than in the GS model. Hence, the brane source enhances the modes near the core.

In the GS model, the eigenfunction with a lowest mass resemblances the massless mode with an almost vanishing amplitude, it was referred as transient mode. This transient mode is absent in the string-cigar model what can be regarded as an improvement since the light mode should already have been detected. Then, the resolution of the GS model get rides this transient mode.

The string-cigar model, unlike the GS model, allows the existence of resonant states. Indeed, the conical behavior near the origin yields to a potential well
not presented in the GS model. Resonance peaks were found (where the highest one occurs when $c = 2.9$ and $m = 0.4192$) which decrease becoming broader with $c$, meaning that the resonant massive states
lifetime decreases with the cosmological constant. The solution for the mass eigenvalue indicated with the highest resonance peak shows that this particular massive graviton has a high probability to be found on the brane.
%From this result, we may argue that a KK gravitational mode may exist in our world as a resonant state.

%urthermore, numerical solutions of the Schroedinger-like equation of the  braneworld disclosed massives .

The results obtained points as perspectives the effects of the gravitational KK modes in a modification of the Newtonian potential. In fact,
the higher amplitude of the eigenfunctions at the brane core and the new decreasing behavior of the spectrum may lead to interesting modifications
in the gravitational potential. Further, Figs. \ref{Fig-Ressonancia} and \ref{Fig-Funcao-Onda} show that  the first resonant peak points out the more expressive massive state to contribute to corrections of the four-dimensional gravity laws.

\section*{Acknowledgments}
The authors thank the Funda\c{c}\~{a}o Cearense de apoio ao Desenvolvimento
Cient\'{\i}fico e Tecnol\'{o}gico (FUNCAP), the Coordena\c{c}\~{a}o de Aperfei\c{c}oamento de Pessoal de N\' ivel Superior (CAPES), the Conselho Nacional de Desenvolvimento Cient\' ifico e Tecnol\' ogico (CNPq) and the Instituto Federal do Cear\'a (IFCE) for financial support.

%%%%%%%%%%%%%%%%%%%%%%%%%%%%%%%%%%%%%%%%%%%%%%%%%%%%%%%%%%%%%%%%%%%%%%%%%%%%%%%%%%%%%%%%%%%%%%%%%%%%%%%%%%%%
%%%%%%%%%%%%%%%%%%%%%%%%%%%%%%%%%%%%  REFERENCES %%%%%%%%%%%%%%%%%%%%%%%%%%%%%%%%%%%%%%%%%%%%%%%%%%%%%%%%%%%
%%%%%%%%%%%%%%%%%%%%%%%%%%%%%%%%%%%%%%%%%%%%%%%%%%%%%%%%%%%%%%%%%%%%%%%%%%%%%%%%%%%%%%%%%%%%%%%%%%%%%%%%%%%%

%\newpage

\end{document}